\documentclass[journal]{IEEEtran}
\usepackage[T1]{fontenc}
\usepackage[latin9]{inputenc}
\usepackage{amsthm}
\usepackage{amsmath} 
\usepackage{graphicx} 
\usepackage{color} 
\usepackage{cite}
\usepackage{algpseudocode}
\usepackage{algorithm}
\usepackage{amssymb}
\usepackage{subfigure}
\usepackage{esint}
\usepackage{algpseudocode}
\usepackage{epstopdf}
\usepackage{multirow}
\usepackage{makecell}
\usepackage{float}
\usepackage{lipsum}
\usepackage{balance}

\newtheorem{definition}{Definition}
\newtheorem{remark}{Remark}

\newtheorem{example}{Example}

\theoremstyle{plain}
\theoremstyle{plain}
\newtheorem{theorem}{Theorem}
\newtheorem{lemma}{Lemma}

\newcommand{\comment}[1]{}

\IEEEoverridecommandlockouts

\begin{document}

\title{LOCO Codes: Lexicographically-Ordered Constrained Codes}

\author{
   \IEEEauthorblockN{Ahmed~Hareedy,~\IEEEmembership{Member,~IEEE}, and Robert~Calderbank,~\IEEEmembership{Fellow,~IEEE}\vspace{-1.0em}}
   
   \thanks{A. Hareedy and R. Calderbank are with the Department of Electrical and Computer Engineering, Duke University, Durham, NC 27705 USA (e-mail: ahmed.hareedy@duke.edu; robert.calderbank@duke.edu). This research was supported in part by NSF under grant CCF 1717602. Part of the paper was presented at IEEE Information Theory Workshop (ITW), 2019 \cite{ahh_locoitw}.}
}
\maketitle

%%%%%%%%%%%%%%%%%%%%%%%%%%%%%%%%%%%%%
\begin{abstract}
Line codes make it possible to mitigate interference, to prevent short pulses, and to generate streams of bipolar signals with no direct-current (DC) power content through balancing. They find application in magnetic recording (MR) devices, in Flash devices, in optical recording devices, and in some computer standards. This paper introduces a new family of fixed-length, binary constrained codes, named lexicographically-ordered constrained codes (LOCO codes), for bipolar non-return-to-zero signaling. LOCO codes are capacity-achieving, the lexicographic indexing enables simple, practical encoding and decoding, and this simplicity is demonstrated through analysis of circuit complexity. LOCO codes are easy to balance, and their inherent symmetry minimizes the rate loss with respect to unbalanced codes having the same constraints. Furthermore, LOCO codes that forbid certain patterns can be used to alleviate inter-symbol interference in MR systems and inter-cell interference in Flash systems. Numerical results demonstrate a gain of up to 10\% in rate achieved by LOCO codes with respect to other practical constrained codes, including run-length-limited codes, designed for the same purpose. Simulation results suggest that it is possible to achieve a channel density gain of about 20\% in MR systems by using a LOCO code to encode only the parity bits, limiting the rate loss, of a low-density parity-check code before writing.
\end{abstract}


%%%%%%%%%%%%%%%%%%%%%%%%%%%%%%%%%%%%%
\section{Introduction}\label{sec_intro}

From data storage to data transmission, line codes are employed in many systems to achieve a variety of goals. An important early example, introduced in \cite{tang_bahl}, is the family of run-length-limited (RLL) codes used to mitigate inter-symbol interference (ISI) in magnetic recording (MR) systems by appropriately separating transitions. RLL codes are associated with bipolar non-return-to-zero inverted (NRZI) signaling, where a $0$ is represented by no transition and a $1$ is represented by a transition, with the transitions being from $-A$ to $+A$, $A > 0$, and vice versa. RLL codes are characterized by a pair of parameters, $(d, k)$, where $d$ (resp., $k$) is the minimum (resp., maximum) number of $0$'s between adjacent $1$'s. The parameter $d$ separates transitions, and the parameter $k$ supports self-clocking by ensuring frequent transitions. A variable-length fixed-rate $(2, 7)$ RLL code appeared in the IBM 3370, 3375, and 3380 disk drives \cite{siegel_mr}, and the issue of error propagation for $(2, 7)$ RLL codes was studied in \cite{howe_prop}.

For simplicity, we abbreviate a run of $r$ consecutive $0$'s (resp., $1$'s) to $\bold{0}^r$ (resp., $\bold{1}^r$). A $\mathcal{T}_x$-constrained code is a code that forbids the patterns in $\mathcal{T}_x \triangleq \{0 \bold{1}^y 0, 1 \bold{0}^y 1 \text{ } | \text{ } 1 \leq y \leq x\}$ from appearing in any codeword. $\mathcal{T}_x$-constrained codes are associated with bipolar non-return-to-zero (NRZ) signaling, where a $0$ is represented by level $-A$ and a $1$ is represented by level $+A$. The parameter $x$ separates transitions, which mitigates ISI, serving the same purpose as the parameter $d$ in RLL codes. For example, transitions separated by one bit duration can be prevented by a $\{010, 101\}$-constrained code with NRZ signaling, or a $(1, \infty)$ RLL code with NRZI signaling. We focus in this paper on $\mathcal{T}_x$-constrained codes.

Constrained codes were used to extend the life of MR systems employing peak detection, and they continue to be used in modern MR systems \cite{vasic_prc, col_detect} to improve the performance of sequence detection on partial response (PR) channels such as extended PR4 (EPR4 and E$^2$PR4) channels \cite{siegel_const, immink_1}. PR channels with equalization targets that follow the channel impulse response \cite{ahh_bas} require forbidden patterns to be symmetric. Moreover, constrained codes improve the performance on low resolution media by preventing short pulses, which might be missed when reading \cite{harada_resol}. As $x$ for a $\mathcal{T}_x$-constrained code or $d$ for an RLL code increases, the minimum width of a pulse in the stream to be written increases.

The requirement that the power spectrum of a line code vanishes at frequency zero, i.e., the code is direct-current-free (DC-free), is important in optical recording \cite{immink_opt} and in digital communication over transmission lines. This requirement is typically accomplished by balancing signal signs in the stream of transmitted (written) codewords. The author in \cite{knuth_bal} developed a particularly elegant method of achieving balance, which requires the addition of more than $\log_2 m$ bits, where $m$ is the code length, and this method was later tailored to RLL codes \cite{knuth_mod}. The null at DC can be widened by constraining the higher order statistics of line codewords (see \cite{robert_spec1} and \cite{robert_spec2} for a frequency domain approach).

Constrained codes also find application in Flash memories. Consider a single-level cell (SLC) Flash memory system (the SLC nomenclature is inaccurate; it is rather a single-bit cell with two levels). Given three adjacent cells, the pattern $101$ translates to programming the outer two cells but not the inner cell. This pattern can result in inter-cell interference (ICI) caused by an unintentional increase of the charge level in the inner cell. The pattern $010$ is typically less detrimental, but it can cause problems when erasure is not applied to the entire block and the outer cells are initially programmed. See \cite{qin_flash} for a study of balanced constrained codes that alleviate ICI in Flash systems by eliminating the pattern $(q-1)0(q-1)$, where $q$ is the number of levels in the cell and also the Galois field (GF) order\footnote{We directly map the elements of the GF to the consecutive integers $\{0, 1, \allowbreak \dots, q-1\}$ indexing $q$ distinct threshold voltage levels in order to follow the reference. In Flash systems, NRZ signaling is typically adopted.}. Another related work is \cite{kayser_flash}.

Furthermore, line codes find application in computer standards for data transmission, such as universal serial bus (USB) and peripheral component interconnect express (PCIe). Line codes for these applications are simpler than $\mathcal{T}_x$-constrained and RLL codes, since streams of codewords are only required to be balanced and to support self-clocking. Examples include the $8$b/$10$b code \cite{immink_2}, the $64$b/$66$b code \cite{walker_66}, and the $128$b/$132$b code \cite{saade_line}. We note that constrained codes preserving parity are studied in \cite{roth_pres}, and that constrained codes for deoxyribonucleic acid (DNA) storage are studied in \cite{immink_3}. We refer the reader to \cite{immink_1} for a comprehensive survey of constrained codes available until 1998.

The idea of lexicographic indexing can be traced back to \cite{tang_bahl} and to \cite{cover_lex}. The latter independently introduced the idea in the context of source coding. The RLL codes and balanced RLL codes constructed in \cite{immink_lex} and \cite{braun_lex}, respectively, are based on \cite{cover_lex}, and the rates achieved improve upon those of earlier RLL codes. However, these gains are only realized at relatively large code lengths, and therefore at a significant cost in terms of complexity, storage overhead, and error performance. Moreover, the technique in \cite{cover_lex} does not readily generalize to $\mathcal{T}_x$-constrained codes. While techniques based on lookup tables, e.g., \cite{immink_table}, offer a better rate-length trade-off, they incur significant encoding and decoding complexity.

In this paper, we return to the presentation of lexicographic indexing in \cite{tang_bahl}, and develop the idea in the context of a new family of $\mathcal{T}_x$-constrained codes. We call the new codes lexicographically-ordered $\mathcal{T}_x$-constrained codes, or simply LOCO codes. Our three most significant contributions are:

\begin{enumerate}
\item We develop a simple rule for encoding and decoding LOCO codes based on lexicographic indexing. This rule reduces the encoding-decoding of LOCO codes to low-complexity mapping-demapping between the index of a codeword and the codeword itself. We demonstrate that LOCO codes are capacity-achieving codes, and that at moderate lengths, they provide a rate gain of up to $10\%$ compared with practical RLL and other $\mathcal{T}_x$-constrained codes that are used to achieve the same goals.

\item We demonstrate a density gain of about $20\%$ in modern MR systems by using a LOCO code to protect only the parity bits of a low-density parity-check (LDPC) code via alleviating ISI. The density gain of LDPC-LOCO coding compared with same-rate LDPC coding is about $16\%$. It is of course possible to protect all the bits of the LDPC code, but our method limits the rate loss. Our demonstration uses a modified version of the PR system described in \cite{ahh_bas}, and a spatially-coupled (SC) LDPC code constructed as in \cite{ahh_tit2}.

\item We prove that the inherent symmetry of LOCO codes makes balancing easy. Each message in a balanced LOCO code is represented by two codewords that are the complements of each other. Moreover, we show that the rate loss in balancing LOCO codes is minimal, and that this loss tends to zero in the limit, so that balanced LOCO codes achieve the same asymptotic rates as their unbalanced counterparts.
\end{enumerate}

We also describe how to modify LOCO codes to achieve self-clocking with NRZ signaling.

The rest of the paper is organized as follows. In Section~\ref{sec_lex}, LOCO codes are formally defined and analyzed. The mapping-demapping between the index of a codeword and the codeword itself is introduced in Section~\ref{sec_pract}. Next, the rates of LOCO codes in addition to the practical encoding and decoding algorithms are presented in Section~\ref{sec_ralg}. LOCO codes are applied to MR systems in Section~\ref{sec_mr}. Balanced LOCO codes and their rates are discussed in Section~\ref{sec_bala}. Finally, the paper is concluded in Section~\ref{sec_conc}.

%%%%%%%%%%%%%%%%%%%%%%%%%%%%%%%%%%%%%
\section{Analysis of LOCO Codes}\label{sec_lex}

We start with the formal definition of the proposed fixed-length LOCO codes. In the next two sections, we will propose simple, practical encoding and decoding schemes for these codes.

\begin{definition}\label{def_loco}
A LOCO code $\mathcal{C}_{m,x}$, with parameters $m \geq 1$ and $x \geq 1$, is defined by the following properties:

\begin{enumerate}
\item Each codeword $\bold{c} \in \mathcal{C}_{m,x}$ is binary and of length $m$.
\item Codewords in $\mathcal{C}_{m,x}$ are ordered lexicographically.
\item Each codeword $\bold{c} \in \mathcal{C}_{m,x}$ does not contain any pattern in the set $\mathcal{T}_x$, where:
\begin{equation}\label{eqn_Tx}
\mathcal{T}_x \triangleq \{010, 101, 0110, 1001, \dots, 0 \bold{1}^x 0, 1 \bold{0}^x 1\};
\end{equation}
therefore, $\vert \mathcal{T}_x \vert = 2x$.
\item Codewords in $\mathcal{C}_{m,x}$ are all the codewords satisfying the previous three conditions.
\end{enumerate}
\end{definition}

Lexicographic ordering of codewords means that the codewords are ordered in an ascending manner following the rule $0 < 1$ for any bit, and the bit significance reduces from left to right. In particular, starting from the left, we say $\bold{c}_{u_1} < \bold{c}_{u_2}$ if and only if for the first bit position the two codewords differ at, $\bold{c}_{u_1}$ has $0$ while $\bold{c}_{u_2}$ has $1$.

Since $\mathcal{T}_x$-constrained codes are used with NRZ signaling, the constrained set of patterns can also be written as:
\begin{align}
\{&-+\hspace{+0.2em}-, +-+, -++\hspace{+0.2em}-, +--\hspace{+0.2em}+, \dots, - \boldsymbol{+}^x -, + \boldsymbol{-}^x +\}, \nonumber
\end{align}
where the notation $\boldsymbol{-}^r$ (resp., $\boldsymbol{+}^r$) is defined the same way as $\bold{0}^r$ (resp., $\bold{1}^r$). Throughout the paper, NRZ (resp., NRZI) signaling is adopted for LOCO (resp., RLL) codes.

\begin{remark}
In the case of Flash systems, the level $-A$ is replaced by the erasure level $E$, $E < A$.
\end{remark}

Observe the connection between the forbidden patterns, i.e., the patterns in $\mathcal{T}_x$, and the physics of different data storage systems. As $x$ increases, ISI (resp., ICI) is more alleviated in MR (resp., Flash) systems, and the minimum width of a pulse increases. However, increasing $x$ reduces the rate of the LOCO code.

Table~\ref{table_1} presents the LOCO codes $\mathcal{C}_{m,1}$, $m \in \{1, 2, \dots, 6\}$. These LOCO codes have $x=1$ and $\mathcal{T}_1 = \{010, 101\}$.

For $m \geq 2$, we partition the codewords in $\mathcal{C}_{m,x}$ into four distinct groups as follows:

\textbf{Group~1:} Codewords in this group start with $00$ from the left, i.e., at the left-most bits (LMBs).

\textbf{Group~2:} Codewords in this group start with $11$ from the left, i.e., at the LMBs.

\textbf{Group~3:} Codewords in this group start with $1\bold{0}^{x+1}$ from the left, i.e., at the LMBs.

\textbf{Group~4:} Codewords in this group start with $0\bold{1}^{x+1}$ from the left, i.e., at the LMBs\footnote{In Groups~3 and 4 and with $2 \leq m \leq x+1$, there exists only a single codeword, which has fewer bits than these LMBs, in each group. The following analysis also applies for such codewords.}.

The four groups are shown in Table~\ref{table_1} for the code $\mathcal{C}_{6, 1}$.

We will see that this partitioning into groups enables enumeration in addition to low complexity encoding and decoding of LOCO codewords.

\begin{remark}
In order to satisfy Condition~3 in Definition \ref{def_loco} for a stream of codewords of a LOCO code $\mathcal{C}_{m,x}$, a bridging pattern needs to be added between any two consecutively transmitted (written) codewords in this stream. Bridging patterns will be discussed later in this paper.
\end{remark}

\begin{table*}
\caption{All the codewords of six LOCO codes, $\mathcal{C}_{m,1}$, $m \in \{1, 2, \dots, 6\}$. The four different groups of codewords are explicitly illustrated~for the code $\mathcal{C}_{6,1}$.}
\vspace{-0.5em}
\centering
\scalebox{1.00}
{
\begin{tabular}{|c|c|c|c|c|c|c c|}
\hline
\multirow{2}{*}{Codeword index $g(\bold{c})$} & \multicolumn{7}{|c|}{\makecell{Codewords of the code $\mathcal{C}_{m,1}$}} \\
\cline{2-8}
{} & \makecell{$m=1$} & \makecell{$m=2$} & \makecell{$m=3$} & \makecell{$m=4$} & \makecell{$m=5$} & \multicolumn{2}{|c|}{\makecell{$m=6$}} \\
\hline
$0$ & $0$ & $00$ & $000$ & $0000$ & $00000$ & $000000$ & \multirow{8}{*}{Group~1} \\
\cline{1-1}\cline{2-2}\cline{3-3}
$1$ & $1$ & $01$ & $001$ & $0001$ & $00001$ & $000001$ & \\
\cline{1-1}\cline{2-2}\cline{3-3}\cline{4-4}
$2$ &  & $10$ & $011$ & $0011$ & $00011$ & $000011$ & \\
\cline{1-1}\cline{3-3}\cline{4-4}\cline{5-5}
$3$ &  & $11$ & $100$ & $0110$ & $00110$ & $000110$ & \\
\cline{1-1}\cline{3-3}\cline{4-4}
$4$ &  &  & $110$ & $0111$ & $00111$ & $000111$ & \\
\cline{1-1}\cline{5-5}\cline{6-6}
$5$ &  &  & $111$ & $1000$ & $01100$ & $001100$ & \\
\cline{1-1}\cline{4-4}
$6$ &  &  &  & $1001$ & $01110$ & $001110$ & \\
\cline{1-1}\cline{5-5}
$7$ &  &  &  & $1100$ & $01111$ & $001111$ & \\
\cline{1-1}\cline{6-6}\cline{7-8}
$8$ &  &  &  & $1110$ & $10000$ & $011000$ & \multirow{5}{*}{Group~4} \\
\cline{1-1}
$9$ &  &  &  & $1111$ & $10001$ & $011001$ & \\
\cline{1-1}\cline{5-5}
$10$ &  &  &  &  & $10011$ & $011100$ & \\
\cline{1-1}\cline{6-6}
$11$ &  &  &  &  & $11000$ & $011110$ & \\
\cline{1-1}
$12$ &  &  &  &  & $11001$ & $011111$ & \\
\cline{1-1}\cline{7-8}
$13$ &  &  &  &  & $11100$ & $100000$ & \multirow{5}{*}{Group~3} \\
\cline{1-1}
$14$ &  &  &  &  & $11110$ & $100001$ & \\
\cline{1-1}
$15$ &  &  &  &  & $11111$ & $100011$ & \\
\cline{1-1}\cline{6-6}
$16$ &  &  &  &  &  & $100110$ & \\
\cline{1-1}
$17$ &  &  &  &  &  & $100111$ & \\
\cline{1-1}\cline{7-8}
$18$ &  &  &  &  &  & $110000$ & \multirow{8}{*}{Group~2} \\
\cline{1-1}
$19$ &  &  &  &  &  & $110001$ & \\
\cline{1-1}
$20$ &  &  &  &  &  & $110011$ & \\
\cline{1-1}
$21$ &  &  &  &  &  & $111000$ & \\
\cline{1-1}
$22$ &  &  &  &  &  & $111001$ & \\
\cline{1-1}
$23$ &  &  &  &  &  & $111100$ & \\
\cline{1-1}
$24$ &  &  &  &  &  & $111110$ & \\
\cline{1-1}
$25$ &  &  &  &  &  & $111111$ & \\
\hline
Code cardinality & $N(1, 1) \triangleq 2$  & $N(2, 1) = 4$ & $N(3, 1) = 6$ & $N(4, 1) = 10$ & $N(5, 1) = 16$ &  \multicolumn{2}{|c|}{$N(6, 1) = 26$} \\
\hline
\end{tabular}}
\label{table_1}
\end{table*}

First, we determine the cardinality of $\mathcal{C}_{m,x}$.

\begin{theorem}\label{thm_loco_card}
Let $N(m, x)$ be the cardinality (size) of the LOCO code $\mathcal{C}_{m,x}$, i.e., $N(m, x) = \vert \mathcal{C}_{m,x} \vert$. Define:
\begin{equation}\label{eqn_Ndef}
N(m, x) \triangleq 2, \textup{ } m \leq 1.
\end{equation}
Then, the following recursive formula gives $N(m, x)$:
\begin{equation}\label{eqn_rec}
N(m, x) = N(m-1, x) + N(m-x-1, x), \textup{ } m \geq 2.
\end{equation}
\end{theorem}

\begin{IEEEproof}
Observe first that symmetry of forbidden patterns implies that in $\mathcal{C}_{m,x}$, the number of codewords starting with $0$ from the left, i.e., at the LMB, equals the number of codewords starting with $1$ from the left. Thus, to prove our recursive formula (\ref{eqn_rec}), we calculate the cardinalities of Group~1 and Group~4 in $\mathcal{C}_{m,x}$, $m \geq 2$, then add these cardinalities and multiply the result by $2$.

\textbf{Group~1:} Each codeword in Group~1 in $\mathcal{C}_{m,x}$ corresponds to a codeword in $\mathcal{C}_{m-1,x}$ that starts with $0$ from the left and shares the remaining $m-2$ right-most bits (RMBs) with the codeword in $\mathcal{C}_{m,x}$. This correspondence is bijective. Thus, the cardinality of Group~1 is:
\begin{equation}\label{eqn_card1}
N_1(m, x) = \frac{1}{2} N(m-1, x).
\end{equation}

\textbf{Group~4:} Each codeword in Group~4 in $\mathcal{C}_{m,x}$ corresponds to a codeword in $\mathcal{C}_{m-x-1,x}$ that starts with $1$ from the left and shares the remaining $m-x-2$ RMBs with the codeword in $\mathcal{C}_{m,x}$. This correspondence is bijective. Thus, the cardinality of Group~4 is:
\begin{equation}\label{eqn_card4}
N_4(m, x) = \frac{1}{2} N(m-x-1, x).
\end{equation}

From (\ref{eqn_card1}) and (\ref{eqn_card4}), we get:
\begin{align}
N(m, x) &= 2 \left [ N_1(m, x) + N_4(m, x) \right ] \nonumber \\ &= N(m-1, x) + N(m-x-1, x), \nonumber
\end{align}
which completes the proof.
\end{IEEEproof}

In a similar way, it can be shown that the cardinality of Group~2 is:
\begin{equation}\label{eqn_card2}
N_2(m, x) = \frac{1}{2} N(m-1, x),
\end{equation}
and the cardinality of Group~3 is:
\begin{equation}\label{eqn_card3}
N_3(m, x) = \frac{1}{2} N(m-x-1, x).
\end{equation}

The value of Theorem~\ref{thm_loco_card} is the insight it provides into the structure of $\mathcal{C}_{m,x}$. Not only does Theorem~\ref{thm_loco_card} perform enumeration via simple recursion, it also significantly contributes to the low-complexity encoding and decoding schemes, which are based on the lexicographic ordering. Note that $N(m, x)$ is always even.

For $x=1$, the cardinalities form a Fibonacci sequence as (\ref{eqn_rec}) becomes:
\begin{equation}\label{eqn_recf}
N(m, 1) = N(m-1, 1) + N(m-2, 1).
\end{equation}
The cardinalities $N(m, 1)$ for $m \in \{1, 2, \dots, 6\}$ are given in the last row of Table~\ref{table_1}.

\begin{example}\label{ex_1}
Consider the LOCO codes $\mathcal{C}_{m,1}$, $m \in \{1, 2, \allowbreak \dots, 6\}$, illustrated in Table~\ref{table_1}. From (\ref{eqn_Ndef}), $N(0, 1) \triangleq 2$ and $N(1, 1) \triangleq 2$. From (\ref{eqn_rec}), which is (\ref{eqn_recf}) for $x=1$, the cardinalities of $\mathcal{C}_{m,1}$, $m \in \{2, 3, \dots, 6\}$, are:
\begin{align}
N(2, 1) &= N(1, 1) + N(0, 1) = 2 + 2 = 4, \nonumber \\
N(3, 1) &= N(2, 1) + N(1, 1) = 4 + 2 = 6, \nonumber \\
N(4, 1) &= N(3, 1) + N(2, 1) = 6 + 4 = 10, \nonumber \\
N(5, 1) &= N(4, 1) + N(3, 1) = 10 + 6 = 16, \nonumber \\
N(6, 1) &= N(5, 1) + N(4, 1) = 16 + 10 = 26. \nonumber
\end{align}
The cardinality of $\mathcal{C}_{6,1}$, for example, can also be obtained from the cardinalities of its groups that are:
\begin{align}
N_1(6, 1) &= \frac{1}{2}N(5, 1) = 8, \nonumber \\
N_2(6, 1) &= \frac{1}{2}N(5, 1) = 8, \nonumber \\
N_3(6, 1) &= \frac{1}{2}N(4, 1) = 5, \nonumber \\
N_4(6, 1) &= \frac{1}{2}N(4, 1) = 5. \nonumber
\end{align}
\end{example}

We now use the group structure of LOCO codes to define a lexicographic indexing of codewords.

Define $g(m, x, \bold{c}) \in \{0, 1, \dots, N(m, x)-1\}$ as the index of a codeword $\bold{c}$ in $\mathcal{C}_{m,x}$, which we also abbreviate to $g(\bold{c})$ when the context is clear. In particular, $g(m, x, \bold{c})$ is the index of the codeword $\bold{c}$ in $\mathcal{C}_{m,x}$ when all the codewords of $\mathcal{C}_{m,x}$ are ordered lexicographically. Since the four groups can be defined for a LOCO code of any length, we define them for $\mathcal{C}_{m+1, x}$. Let $\bold{c}'$ be a codeword in $\mathcal{C}_{m+1, x}$. For Groups~1 and 2 in $\mathcal{C}_{m+1, x}$, let $\bold{c} \in \mathcal{C}_{m,x}$ be the corresponding codeword to $\bold{c}' \in \mathcal{C}_{m+1,x}$ according to the proof of Theorem~\ref{thm_loco_card}, i.e., the $m$ RMBs in $\bold{c}'$ are $\bold{c}$.

We define the \textbf{shift in codeword indices} for Groups~1 and 2 in $\mathcal{C}_{m+1, x}$ as follows:
\begin{equation}\label{eqn_shift}
\zeta_\ell \triangleq g(m+1, x, \bold{c}')-g(m, x, \bold{c}), \textup{ } \ell \in \{1, 2\},
\end{equation}
where $\ell$ is the group index. Observe that this shift is fixed for all the codewords in the same group in $\mathcal{C}_{m+1, x}$.

The following lemma gives the values of the shift for Groups~1 and 2.

\begin{lemma}\label{lem_shifts}
The shift in codeword indices defined in (\ref{eqn_shift}) for Groups~1 and 2 in a LOCO code $\mathcal{C}_{m+1, x}$ is given by:
\begin{align}\label{eqn_svals}
\zeta_\ell = \left\{\begin{matrix}0, \textup{ } &\ell = 1,
\\ N(m-x, x), \textup{ } &\ell = 2.
\end{matrix}\right.
\end{align}
\end{lemma}

\begin{IEEEproof}
We prove (\ref{eqn_svals}) by deriving $\zeta_\ell$ for each of the two groups of codewords in $\mathcal{C}_{m+1, x}$ as follows.

\textbf{Group~1:} Since corresponding codewords in $\mathcal{C}_{m+1,x}$ and in $\mathcal{C}_{m,x}$ have the same index for that group, we get:
\begin{equation}\label{eqn_zeta1}
\zeta_1 = g(m+1, x, \bold{c}')-g(m, x, \bold{c}) = 0.
\end{equation}

\textbf{Group~2:} Group~2 in $\mathcal{C}_{m+1,x}$ comes right after Groups~1, 4, and 3 (see Table~\ref{table_1}). On the other hand, the codewords in $\mathcal{C}_{m,x}$ that correspond to the codewords in Group~2 in $\mathcal{C}_{m+1,x}$ come right after all the codewords that start with $0$ from the left. Consequently, and using (\ref{eqn_card1}), (\ref{eqn_card4}), and (\ref{eqn_card3}):
\begin{align}\label{eqn_zeta2}
\zeta_2 &= g(m+1, x, \bold{c}')-g(m, x, \bold{c}) \nonumber \\ &= N_1(m+1, x) + N_4(m+1, x) \nonumber \\ &+ N_3(m+1,1) - \frac{1}{2}N(m, x) \nonumber \\ &= \frac{1}{2} N(m, x) + \frac{1}{2} N(m-x,x) \nonumber \\ &+ \frac{1}{2}N(m-x, x) - \frac{1}{2}N(m, x) \nonumber \\ &= N(m-x, x).
\end{align}

Noting that (\ref{eqn_zeta1}) and (\ref{eqn_zeta2}) combined are (\ref{eqn_svals}) completes the proof.
\end{IEEEproof}

\begin{example}\label{ex_2}
From (\ref{eqn_svals}), the values of $\zeta_\ell$, $\ell \in \{1, 2\}$, for the LOCO code $\mathcal{C}_{6,1}$ given in the last column of Table~\ref{table_1} are:
\begin{align}
\zeta_1 &= 0, \nonumber \\
\zeta_2 &= N(4, 1) = 10. \nonumber
\end{align}
Note that here $m+1 =6$, i.e., $m=5$, and $x=1$.
\end{example}

%%%%%%%%%%%%%%%%%%%%%%%%%%%%%%%%%%%%%
\section{Practical Encoding and Decoding \\ of LOCO Codes}\label{sec_pract}

In this section, we describe how lexicographic indexing supports simple, practical encoding and decoding of LOCO codes. The following theorem is fundamental to the encoding and decoding algorithms presented in Section \ref{sec_ralg}.

In the following, we define a codeword $\bold{c} \in \mathcal{C}_{m,x}$ as $\bold{c} \triangleq \left [ c_{m-1} \textup{ } c_{m-2} \textup{ } \dots \textup{ } c_0 \right ]$, where $c_i \in \{0, 1\}$, for all $i$. We also define an integer variable $a_i$ for each $c_i$ such that:
\begin{align}\label{eqn_ai}
a_i \triangleq \left\{\begin{matrix}1, \textup{ } &c_i = 1,
\\ 0, \textup{ } &c_i = 0.
\end{matrix}\right.
\end{align}
The same notation applies for $\bold{c}' \in \mathcal{C}_{m+1,x}$. Note that codeword indexing is trivial for the case of $m=1$.

\begin{theorem}\label{thm_induct}
Consider a LOCO code $\mathcal{C}_{m,x}$ with $m \geq 2$.~The index $g(\bold{c})$ of a codeword $\bold{c} \in \mathcal{C}_{m,x}$ is derived from $\bold{c}$ itself according to the following equation:
\begin{equation}\label{eqn_g_all}
g(\bold{c}) = \frac{1}{2} \left [ a_{m-1} N(m, x) + \sum_{i=0}^{m-2} a_i N(i-x+1, x) \right ].
\end{equation}
Here, we use the abbreviated notation $g(\bold{c})$ for simplicity.
\end{theorem}

\begin{IEEEproof}
We prove Theorem~\ref{thm_induct} by induction as follows.

\textbf{Base:} The base case here is the case of $m=2$. Let the four available codewords in $\mathcal{C}_{2,x}$ be $\bold{c}_0$, $\bold{c}_1$, $\bold{c}_2$, and $\bold{c}_3$, with the subscript of $\bold{c}$ being its index. The four codewords are shown in Table~\ref{table_1}. The bits of codeword $\bold{c}_u$ are $c_{u,i}$, $i \in \{0, 1\}$, and $a_{u,i}$ is defined for each $c_{u,i}$ as in (\ref{eqn_ai}). We need to prove that (\ref{eqn_g_all}) yields $g(\bold{c}_u) = u$, $u \in \{0, 1, 2, 3\}$.
\begin{align}
g(\bold{c}_0) &= \frac{1}{2} \left [ 0 + \sum_{i=0}^{0} a_{0,i} N(i-x+1, x) \right ] = 0, \nonumber \\ g(\bold{c}_1) &= \frac{1}{2} \left [ 0 + \sum_{i=0}^{0} a_{1,i} N(i-x+1, x) \right ] \nonumber \\ &= \frac{1}{2} N(-x+1, x) = 1, \nonumber \\ g(\bold{c}_2) &= \frac{1}{2} \left [ N(2, x) + \sum_{i=0}^{0} a_{2,i} N(i-x+1, x) \right ] \nonumber \\ &= \frac{1}{2} \left [ 4 + 0 \right ] = 2, \nonumber \\ g(\bold{c}_3) &= \frac{1}{2} \left [ N(2, x) + \sum_{i=0}^{0} a_{3,i} N(i-x+1, x) \right ] \nonumber \\ &= \frac{1}{2} \left [ 4 + N(-x+1, x)  \right ] = \frac{1}{2} \left [ 4 + 2 \right ] = 3.
\end{align}
Note that $N(-x+1, x) = 2$, for all $x \in \{1, 2, \dots\}$, follows directly from (\ref{eqn_Ndef}). Note also that $N(2, x) = 4$, for all $x \in \{1, 2, \dots\}$.

\textbf{Assumption:} We assume that (\ref{eqn_g_all}) holds for the case of $\overline{m} \in \{2, 3, \dots, m\}$, i.e., for all the LOCO codes $\mathcal{C}_{\overline{m},x}$ of length $\overline{m} \in \{2, 3, \dots, m\}$. In particular,
\begin{equation}\label{eqn_asum}
g(\overline{m}, x, \overline{\bold{c}}) = \frac{1}{2} \left [ \overline{a}_{\overline{m}-1} N(\overline{m}, x) + \sum_{i=0}^{\overline{m}-2} \overline{a}_i N(i-x+1, x) \right ].
\end{equation}
Note that $\overline{\bold{c}}$ with bits $\overline{c}_i$ and variables $\overline{a}_i$, $i \in \{0, 1, \dots, \overline{m}-1\}$, is a codeword in the LOCO code $\mathcal{C}_{\overline{m},x}$.

\textbf{To be proved:} We prove that (\ref{eqn_g_all}) holds for the case of $m+1$, i.e., for the LOCO code $\mathcal{C}_{m+1,x}$ of length $m+1$. In particular,
\begin{align}\label{eqn_prove}
&g(m+1, x, \bold{c}') \nonumber \\ &= \frac{1}{2} \left [ a'_{m} N(m+1, x) + \sum_{i=0}^{m-1} a'_i N(i-x+1, x) \right ].
\end{align}

We prove (\ref{eqn_prove}) for the four groups of codewords in $\mathcal{C}_{m+1,x}$, making use of the inductive assumption and Lemma~\ref{lem_shifts}.

\textbf{Group~1:} From (\ref{eqn_zeta1}), we know that for Group~1:
\begin{equation}
g(m+1, x, \bold{c}') = g(m, x, \bold{c}). \nonumber
\end{equation}
Note that here $\bold{c}$ starts with $0$ from the left. Consequently, and using the assumption in (\ref{eqn_asum}):
\begin{equation}\label{eqn_temp1}
g(m+1, x, \bold{c}') = \frac{1}{2} \left [ 0 + \sum_{i=0}^{m-2} a_i N(i-x+1, x) \right ].
\end{equation}
Since $\bold{c}'$ and $\bold{c}$ share the $m-1$ RMBs, and since $\bold{c}'$ starts with $00$ from the left, i.e., $a'_m = a'_{m-1} = 0$, (\ref{eqn_temp1}) can be written as:
\begin{align}\label{eqn_res1}
&g(m+1, x, \bold{c}') \nonumber \\ &= \frac{1}{2} \left [ a'_{m} N(m+1, x) + \sum_{i=0}^{m-1} a'_i N(i-x+1, x) \right ].
\end{align}

\textbf{Group~2:} From (\ref{eqn_zeta2}), we know that for Group~2:
\begin{equation}
g(m+1, x, \bold{c}') = g(m, x, \bold{c}) + N(m-x, x). \nonumber
\end{equation}
Note that here $\bold{c}$ starts with $1$ from the left. Consequently, and using the assumption in (\ref{eqn_asum}):
\begin{align}\label{eqn_temp2}
g(m+1, x, \bold{c}') &= \frac{1}{2} \left [ N(m, x) + \sum_{i=0}^{m-2} a_i N(i-x+1, x) \right ] \nonumber \\ & + N(m-x, x).
\end{align}

Observe that using (\ref{eqn_rec}), we have:
\begin{align}\label{eqn_temp6}
&\frac{1}{2} N(m, x) + N(m-x, x) \nonumber \\ &= \frac{1}{2} \left [ N(m, x) + N(m-x, x) + N(m-x, x) \right ] \nonumber \\ &= \frac{1}{2} \left [ N(m+1, x) + N(m-x, x) \right ].
\end{align}

Substituting (\ref{eqn_temp6}) in (\ref{eqn_temp2}) gives:
\begin{align}\label{eqn_temp7}
g(m+1, x, \bold{c}') &= \frac{1}{2} \Bigg [ N(m+1, x) + N(m-x, x) \nonumber \\ &\hspace{+1.0em} + \sum_{i=0}^{m-2} a_i N(i-x+1, x) \Bigg ].
\end{align}
Since $\bold{c}'$ and $\bold{c}$ share the $m-1$ RMBs, and since $\bold{c}'$ starts with $11$ from the left, i.e., $a'_m = a'_{m-1} = 1$, (\ref{eqn_temp7}) can be written as:
\begin{align}\label{eqn_res2}
&g(m+1, x, \bold{c}') \nonumber \\ &= \frac{1}{2} \left [ a'_m N(m+1, x) + \sum_{i=0}^{m-1} a'_i N(i-x+1, x) \right ].
\end{align}

\textbf{Group~3:} Observe that the codewords in Group~3 in $\mathcal{C}_{m+1,x}$ are the first $N_3(m+1, x)$ codewords in Group~1 in $\mathcal{C}_{m+1,x}$ after replacing the $0$ at the LMB with $1$ for each (with the same order). Therefore, to get the index $g(m+1, x, \bold{c}')$ for a codeword in Group~3, we need to add $\frac{1}{2}N(m+1, x)$ to the index of the corresponding codeword in Group~1. Thus, and using (\ref{eqn_res1}), for Group~3:
\begin{align}\label{eqn_temp8}
g(m+1, x, \bold{c}') &= \frac{1}{2} \left [ 0 + \sum_{i=0}^{m-1} a'_i N(i-x+1, x) \right ] \nonumber \\ &+ \frac{1}{2} N(m+1, x) .
\end{align}
Since $\bold{c}'$ starts with $1$ from the left, i.e., $a'_m = 1$, (\ref{eqn_temp8}) can be written as:
\begin{align}\label{eqn_res3}
&g(m+1, x, \bold{c}') \nonumber \\ &= \frac{1}{2} \left [ a'_m N(m+1, x) + \sum_{i=0}^{m-1} a'_i N(i-x+1, x) \right ].
\end{align}

\textbf{Group~4:} Observe that the codewords in Group~4 in $\mathcal{C}_{m+1,x}$ are the last $N_4(m+1, x)$ codewords in Group~2 in $\mathcal{C}_{m+1,x}$ after replacing the $1$ at the LMB with $0$ for each (with the same order). Therefore, to get the index $g(m+1, x, \bold{c}')$ for a codeword in Group~4, we need to subtract $\frac{1}{2}N(m+1, x)$ from the index of the corresponding codeword in Group~2. Thus, and using (\ref{eqn_res2}), for Group~4:
\begin{align}\label{eqn_temp9}
g(m+1, x, \bold{c}') &= \frac{1}{2} \left [ N(m+1, x) + \sum_{i=0}^{m-1} a'_i N(i-x+1, x) \right ] \nonumber \\ &- \frac{1}{2} N(m+1, x) .
\end{align}
Since $\bold{c}'$ starts with $0$ from the left, i.e., $a'_m = 0$, (\ref{eqn_temp9}) can be written as:
\begin{align}\label{eqn_res4}
&g(m+1, x, \bold{c}')  \nonumber \\ &= \frac{1}{2} \left [ a'_m N(m+1, x) + \sum_{i=0}^{m-1} a'_i N(i-x+1, x) \right ].
\end{align}

As a result of the above analysis for the four groups, (\ref{eqn_prove}) is proved, i.e., the induction is proved. Therefore, Theorem~\ref{thm_induct} is proved for any LOCO code $\mathcal{C}_{m,x}$, for all $m \geq 2$ and for all $x \geq 1$.
\end{IEEEproof}

Observe that from Theorem~\ref{thm_induct}, two LOCO codewords that differ only in the bit $c_i$, $0 \leq i \leq m-2$, satisfy the following:
\begin{align}\label{eqn_inter}
&g(\left [ c_{m-1} \textup{ } \dots \textup{ } c_{i+1} \textup{ } 1 \textup{ } c_{i-1} \textup{ } \dots \textup{ } c_0 \right ]) - \nonumber \\ &g(\left [ c_{m-1} \textup{ } \dots \textup{ } c_{i+1} \textup{ } 0 \textup{ } c_{i-1} \textup{ } \dots \textup{ } c_0 \right ]) = \frac{1}{2} N(i-x+1, x).
\end{align}
For simplicity, consider the case of $x \leq i \leq m-2$. In order that these two LOCO codewords exist, if $c_{i+1} = 0$, $\left [ c_{i-1} \textup{ } c_{i-2} \textup{ } \dots \textup{ } c_{i-x} \right ]$ is guaranteed to be $\bold{1}^x$, and if $c_{i+1} = 1$, $\left [ c_{i-1} \textup{ } c_{i-2} \textup{ } \dots \textup{ } c_{i-x} \right ]$ is guaranteed to be $\bold{0}^x$. Consequently, the interpretation of (\ref{eqn_inter}) is that this difference in indices equals exactly the number of LOCO codewords of length $i-x+1$ that start with $1$ (resp., $0$) from the left if $c_{i+1} = 0$ (resp., $c_{i+1} = 1$). In both cases, this number is $\frac{1}{2} N(i-x+1, x)$.

The value of Theorem~\ref{thm_induct} is that it provides the mathematical foundation for the practical encoding and decoding algorithms of our LOCO codes via lexicographic indexing. In particular, this theorem introduces a simple one-to-one mapping from $g(\bold{c})$ to $\bold{c}$, which is actually the encoding, and a simple one-to-one demapping from $\bold{c}$ to $g(\bold{c})$, which is actually the decoding. The value of this theorem is exemplified in the practical algorithms in the following section. In summary, Theorem~\ref{thm_induct} provides the encoding-decoding rule for LOCO codes.

\begin{example}\label{ex_3}
We illustrate Theorem~\ref{thm_induct} by applying (\ref{eqn_g_all}) to two codewords in $\mathcal{C}_{6,1}$ given in Table~\ref{table_1}. The first codeword is the one with the index $9$, which is $011001$. This codeword has $c_{m-1}=0$; thus,
\begin{align}
g(\bold{c}) &= \frac{1}{2} \left [ 0 + \sum_{i=0}^{4} a_i N(i, 1) \right ] \nonumber \\ &= \frac{1}{2} \left [ N(4, 1) + N(3, 1) + N(0, 1) \right ] \nonumber \\ &= \frac{1}{2} \left [ 10 + 6 + 2 \right ] = 9. \nonumber
\end{align}
The second codeword is the one with the index $24$, which is $111110$. This codeword has $c_{m-1}=1$; thus,
\begin{align}
g(\bold{c}) &= \frac{1}{2} \left [ N(6, 1) + \sum_{i=0}^{4} a_i N(i, 1) \right ] \nonumber \\ &= \frac{1}{2} \left [ 26 + N(4, 1) + N(3, 1) + N(2, 1) + N(1, 1)  \right ] \nonumber \\ &= \frac{1}{2} \left [ 26 + 10 + 6 + 4 + 2  \right ] = 24. \nonumber
\end{align}
\end{example}

Example~\ref{ex_3} shows how the index, which implies the original message, can be recovered from the LOCO codeword.

\begin{remark}\label{rmk_lorll}
Lexicographically-ordered RLL (LO-RLL) codes can be constructed as shown in \cite{tang_bahl}. Define the binary difference vector $\bold{v}$ of a codeword $\bold{c}$ in a LOCO code $\mathcal{C}_{m,x}$, $m \geq 2$, as $\bold{v} \triangleq \left [ v_{m-2} \textup{ } v_{m-3} \textup{ } \dots \textup{ } v_0 \right ]$, with $v_i \triangleq c_{i+1} + c_i$ over GF($2$), for all $i \in \{0, 1, \dots, m-2\}$. Observe that any codeword $\bold{c}$ of length $m$ in $\mathcal{C}_{m,x}$ has its difference vector $\bold{v}$ of length $m-1$ satisfying the $(d, \infty)$, $d=x$, RLL constraint. Thus, all the codewords of a $(d, \infty)$ LO-RLL code with $d=x$ and length $m-1$ can also be derived from the LOCO code $\mathcal{C}_{m,x}$ by computing the difference vectors for all the codewords in $\mathcal{C}_{m,x}$ starting with $0$ from the left (the remaining difference vectors will be repeated because of symmetry)\footnote{Even though codewords here are not ordered lexicographically, we still call this code a LO-RLL code since all the codewords satisfying the constraint are included and the generating codewords are ordered lexicographically.}. Consequently, the cardinality of a $(d, \infty)$ LO-RLL code with $d=x$ and length $m-1$ is given by:
\begin{equation}\label{eqn_lorll_loco}
N_{\textup{RLL}}(m-1, d) = \frac{1}{2} N(m, x), \textup{ } d=x.
\end{equation}

From \cite{tang_bahl}, the cardinality of a $(d, \infty)$ LO-RLL code of length $m$ is given by:
\begin{equation}\label{eqn_lorll_1}
N_{\textup{RLL}}(m, d) \triangleq 1, \textup{ } m \leq 0, \text{ and}
\end{equation}
\begin{equation}\label{eqn_lorll_2}
N_{\textup{RLL}}(m, d) = N_{\textup{RLL}}(m-1, d) + N_{\textup{RLL}}(m-d-1, d), \textup{ } m \geq 1.
\end{equation}
Comparing (\ref{eqn_lorll_1}) and (\ref{eqn_lorll_2}) to (\ref{eqn_Ndef}) and (\ref{eqn_rec}) results in (\ref{eqn_lorll_loco}). For example, for $x=1$, $N(1, 1) \triangleq 2$, $N(2, 1) = 4$, $N(3, 1) = 6$, $N(4, 1) = 10$, $N(5, 1) = 16$, $N(6, 1) = 26$, \dots. On the other hand, for $d=x=1$, $N_{\textup{RLL}}(1, 1) = 2$, $N_{\textup{RLL}}(2, 1) = 3$, $N_{\textup{RLL}}(3, 1) = 5$, $N_{\textup{RLL}}(4, 1) = 8$, $N_{\textup{RLL}}(5, 1) = 13$, \dots, which demonstrates (\ref{eqn_lorll_loco}). This observation leads to a simple way of constructing and indexing $(d, \infty)$ RLL codes.
\end{remark}

%%%%%%%%%%%%%%%%%%%%%%%%%%%%%%%%%%%%%
\section{Rate Discussion and Algorithms}\label{sec_ralg}

We first discuss \textbf{bridging patterns}. Consider the following scenario. The codeword at transmission (writing) instance $t$ is ending with $00$ from the right, while the codeword at instance $t+1$ is starting with $10$ from the left. The stream containing the two codewords will then have the pattern $010$, which is a forbidden pattern for any LOCO code. This is the motivation behind adding bridging patterns. In particular, bridging patterns prevent forbidden patterns from appearing across two consecutive codewords. If the patterns in $\mathcal{T}_x$ are prevented (Condition~3 in Definition \ref{def_loco} is satisfied), any two consecutive transitions will be separated by at least $x+1$ successive bit durations. For $\mathcal{T}_x$-constrained codes, since they are associated with NRZ signaling, transitions are either from $0$ to $1$, i.e, $-A$ to $+A$, or from $1$ to $0$, i.e., $+A$ to $-A$.

Define the symbol $z$ as the no transmission (no writing) symbol. For example, in magnetic recording, $z$ represents the state when the magnetic grain is unmagnetized. As done before, we also define the notation $\bold{z}^r$ to represent a run of $r$ consecutive $z$ symbols. We propose two methods for adding bridging patterns that prevent forbidden patterns from appearing in streams of LOCO codewords. The first method is simply to add the bridging pattern $\bold{z}^x$ between each two consecutive LOCO codewords. The second method is to make a run-time decision on the bridging pattern of length $x$ based on the $x+1$ RMBs in the codeword at instance $t$ and the $x+1$ LMBs in the codeword at instance $t+1$.

In the first method, adding a run of $x$ consecutive $z$~symbols, i.e., not transmitting (not writing) for $x$ successive bit durations, guarantees that no pattern in $\mathcal{T}_x$ appears across consecutive LOCO codewords in $\mathcal{C}_{m, x}$. This method is quite simple, and does not require any knowledge of the codewords being transmitted (written). However, it is not optimal in the sense that it does not provide the maximum achievable protection, e.g., from ISI in MR systems, for the bits at the two ends of the codeword. For example, in the scenario at the start of this section, it is best to use $1$ for bridging if $x=1$.

While the second method provides better protection for the bits at the two ends of the codeword, it introduces additional complexity and latency. However, it is still feasible for small values of $x$. For example, Table \ref{table_2} provides the bridging patterns of the second method for LOCO codes with $x=1$.

\begin{table}
\caption{Bridging patterns of the second method for LOCO codes with $x=1$.}
\vspace{-0.5em}
\centering
\scalebox{1.00}
{
\begin{tabular}{|c|c|c|}
\hline
\makecell{RMB(s) \\ at instance $t$} & \makecell{Bridging \\ pattern} & \makecell{LMB(s) \\ at instance \\ $t+1$} \\
\hline
$0$ & $0$ & $0$ \\
\hline
$0$ & $0$ & $11$ \\
\hline
$00$ & $1$ & $10$ \\
\hline
$01$ & $z$ & $01$ \\
\hline
$10$ & $z$ & $10$ \\
\hline
$11$ & $0$ & $01$ \\
\hline
$1$ & $1$ & $00$ \\
\hline
$1$ & $1$ & $1$ \\
\hline
\end{tabular}}
\label{table_2}
\vspace{-0.5em}
\end{table}

Whether the first or the second method is used for bridging, the number of added bits/symbols for each codeword is $x$. Moreover, bridging patterns are ignored at the decoding.

\begin{remark}
In the case of Flash systems, transitions are either from $0$ to $1$, i.e, $E$ to $+A$, or from $1$ to $0$, i.e., $+A$ to $E$. Moreover, the no writing symbol $z$ represents the state when the cell is programmed to a charge level about the mid-point between $E$ and $+A$.
\end{remark}

\begin{remark}
For LOCO codes with parameter $x$, the optimal bridging, in terms of bits protection, is different from the second bridging method. In particular, if the RMB of the codeword at instance $t$ is $0$ (resp., $1$), $\bold{0}^x$ (resp., $\bold{1}^x$) is added for bridging after that $0$ (resp., $1$). Moreover, if the LMB of the codeword at instance $t+1$ is $0$ (resp., $1$), $\bold{0}^x$ (resp., $\bold{1}^x$) is added for bridging before that $0$ (resp., $1$). Thus, for this optimal bridging, $2x$ bridging bits are needed, which also keeps the code length fixed. However, such bridging is not efficient in terms of the added redundancy, in addition to its higher complexity compared with the first bridging method. Furthermore, our simulations demonstrate that the other two bridging methods described above are already guaranteeing a more than satisfactory performance.
\end{remark}

One of the important requirements not only in constrained codes, but also in all types of line codes is \textbf{self-clocking} \cite{siegel_mr, immink_1}. In particular, the receiver should be capable of retrieving the clock of the transmitter from the signal itself. This requires avoiding long runs of $0$'s ($-A$'s) and long runs of $1$'s ($+A$'s). To achieve this goal, we construct the following code.

\begin{definition}\label{def_clo}
A self-clocked LOCO (C-LOCO) code $\mathcal{C}_{m,x}^{\textup{c}}$ is the code resulting from removing the all $0$'s and the all $1$'s codewords from the LOCO code $\mathcal{C}_{m,x}$. In particular,
\begin{equation}\label{eqn_scl}
\mathcal{C}_{m,x}^{\textup{c}} \triangleq \mathcal{C}_{m,x} \setminus \{\bold{0}^m,  \bold{1}^m\},
\end{equation}
where $m \geq 2$. The cardinality of $\mathcal{C}_{m,x}^{\textup{c}}$ is given by:
\begin{equation}\label{eqn_scl_card}
N^{\textup{c}}(m, x) = N(m, x) - 2.
\end{equation}
\end{definition}

Now, there exists at least one transition in each codeword in $\mathcal{C}_{m,x}^{\textup{c}}$. Define $k_{\textup{eff}}^{\textup{c}}$ as the maximum number of successive bit durations between two consecutive transitions in a stream of C-LOCO codewords that belong to $\mathcal{C}_{m,x}^{\textup{c}}$, with each two consecutive codewords separated by a bridging pattern. For the sake of abbreviation, we here use the format ``codeword at $t$ $-$ bridging pattern $-$ codeword at $t+1$''. The scenarios under which $k_{\textup{eff}}^{\textup{c}}$ is achieved, using the first bridging method, are:
\begin{align}
&1\bold{0}^{m-1} - \bold{z}^x - \bold{0}^{m-1}1 \textup{ } \textup{ and } \nonumber \\
&0\bold{1}^{m-1} - \bold{z}^x - \bold{1}^{m-1}0. \nonumber
\end{align}
The scenarios under which $k_{\textup{eff}}^{\textup{c}}$ is achieved, using the second bridging method, are:
\begin{align}
&1\bold{0}^{m-1} - \bold{0}^x - \bold{0}^{m-1}1 \textup{ } \textup{ and } \nonumber \\
&0\bold{1}^{m-1} - \bold{1}^x - \bold{1}^{m-1}0. \nonumber
\end{align}
Observe that a transition is only from $0$ to $1$ or from $1$ to $0$. Consequently, regardless of the chosen method, we get:
\begin{equation}\label{eqn_keff}
k_{\textup{eff}}^{\textup{c}} = 2(m-1) + x.
\end{equation}

We are now ready to discuss the rate of C-LOCO codes. A C-LOCO code $\mathcal{C}_{m,x}^{\textup{c}}$, with $x$ bridging bits/symbols associated to each codeword, has rate:
\begin{align}\label{eqn_rate}
R_{\textup{LOCO}}^{\textup{c}} &= \frac{\left \lfloor \log_2 N^{\textup{c}}(m, x)  \right \rfloor}{m+x} \nonumber \\ &= \frac{\left \lfloor \log_2 \left( N(m, x) - 2 \right )  \right \rfloor}{m+x},
\end{align}
where $N(m, x)$ is obtained from the recursive relation (\ref{eqn_rec}). The numerator, which is $\left \lfloor \log_2 \left( N(m, x) - 2 \right )  \right \rfloor$, is the length of the messages $\mathcal{C}_{m,x}^{\textup{c}}$ encodes.

Observe that a C-LOCO code $\mathcal{C}_{m,x}^{\textup{c}}$ consists of all codewords of length $m$, with the exception of the two codewords $\bold{0}^m$ and $\bold{1}^m$, that do not contain any of the forbidden patterns in $\mathcal{T}_x$. Moreover, the number of added bits/symbols for bridging is function of $x$ only, and thus does not grow with $m$. Consequently, it follows that C-LOCO codes are \textbf{capacity-achieving constrained codes}.

\begin{example}\label{ex_4}
Consider again the LOCO code $\mathcal{C}_{6,1}$ in Table~\ref{table_1}. From (\ref{eqn_keff}), the C-LOCO code $\mathcal{C}_{6,1}^{\textup{c}}$ derived from $\mathcal{C}_{6,1}$ has:
\begin{equation}
k_{\textup{eff}}^{\textup{c}} = 2(6-1) + 1 = 11. \nonumber
\end{equation}

\begin{table}
\caption{The C-LOCO code $\mathcal{C}_{6,1}^{\textup{c}}$ for all messages.}
\vspace{-0.5em}
\centering
\scalebox{1.00}
{
\begin{tabular}{|c|c|c|}
\hline
\makecell{Message} & \makecell{$g(\bold{c})$} & \makecell{Codeword $\bold{c}$} \\
\hline
$0000$ & $1$ & $000001$ \\
\hline
$0001$ & $2$ & $000011$ \\
\hline
$0010$ & $3$ & $000110$ \\
\hline
$0011$ & $4$ & $000111$ \\
\hline
$0100$ & $5$ & $001100$ \\
\hline
$0101$ & $6$ & $001110$ \\
\hline
$0110$ & $7$ & $001111$ \\
\hline
$0111$ & $8$ & $011000$ \\
\hline
$1000$ & $9$ & $011001$ \\
\hline
$1001$ & $10$ & $011100$ \\
\hline
$1010$ & $11$ & $011110$ \\
\hline
$1011$ & $12$ & $011111$ \\
\hline
$1100$ & $13$ & $100000$ \\
\hline
$1101$ & $14$ & $100001$ \\
\hline
$1110$ & $15$ & $100011$ \\
\hline
$1111$ & $16$ & $100110$ \\
\hline
\end{tabular}}
\label{table_3}
\vspace{-0.3em}
\end{table}

The length of the messages $\mathcal{C}_{6,1}^{\textup{c}}$ encodes is:
\begin{equation}
\left \lfloor \log_2 \left( N(6, 1) - 2 \right )  \right \rfloor = \left \lfloor \log_2 24  \right \rfloor = 4. \nonumber
\end{equation}
The C-LOCO code $\mathcal{C}_{6,1}^{\textup{c}}$ is shown in Table~\ref{table_3} for all messages. From (\ref{eqn_rate}), the rate of $\mathcal{C}_{6,1}^{\textup{c}}$ is:
\begin{equation}
R_{\textup{LOCO}}^{\textup{c}} = \frac{\left \lfloor \log_2 24  \right \rfloor}{6+1} = \frac{4}{7} = 0.5714. \nonumber
\end{equation}
\end{example}

Note that the rate of $\mathcal{C}_{6,1}^{\textup{c}}$ is relatively low because of the small code length, $m=6$, and because of the relatively high number of unused codewords. Table~\ref{table_4} shows the rates of C-LOCO codes $\mathcal{C}_{m,x}^{\textup{c}}$ for different values of $m$ and for $x \in \{1, 2\}$. The rates in Table~\ref{table_4} for C-LOCO codes with $x=1$ are significantly higher than $0.5714$.

Table~\ref{table_4} demonstrates that C-LOCO codes have rates up to $0.6923$ (resp., $0.5484$) for the case of $x=1$ (resp., $x=2$) with moderate code lengths. From the literature, the capacity of a $\mathcal{T}_x$-constrained code with $x=1$ (resp., $x=2$) is $0.6942$ (resp., $0.5515$) \cite{siegel_const, immink_1}. The table shows that the rate of the C-LOCO code $\mathcal{C}_{90,1}^{\textup{c}}$ (resp., $\mathcal{C}_{91,2}^{\textup{c}}$) is within only $0.3\%$ (resp., $0.6\%$) from the capacity. In fact, these rates even increase with an informed increase in the value of $m$ until they reach the capacity. For example, the rate of $\mathcal{C}_{489,1}^{\textup{c}}$ is $0.6939$, which is only $0.04\%$ from the capacity. Additionally, the rate of $\mathcal{C}_{450,2}^{\textup{c}}$ is $0.5509$, which is only $0.1\%$ from the capacity.

\begin{table}
\caption{Rates and adder sizes of C-LOCO codes $\mathcal{C}_{m,x}^{\textup{c}}$ for different values of $m$ and $x$. The capacity is $0.6942$ for $x=1$ and $0.5515$ for $x=2$.}
\vspace{-0.5em}
\centering
\scalebox{1.00}
{
\begin{tabular}{|c|c|c|}
\hline
\makecell{Values of $m$ and $x$} & \makecell{$R_{\textup{LOCO}}^{\textup{c}}$} & \makecell{Adder size} \\
\hline
$m=8$, $x=1$ & $0.6667$ & $6$ bits \\
\hline
$m=18$, $x=1$ & $0.6842$ & $13$ bits \\
\hline
$m=31$, $x=1$ & $0.6875$ & $22$ bits \\
\hline
$m=44$, $x=1$ & $0.6889$ & $31$ bits \\
\hline
$m=54$, $x=1$ & $0.6909$ & $38$ bits \\
\hline
$m=90$, $x=1$ & $0.6923$ & $63$ bits \\
\hline
$m=6$, $x=2$ & $0.5000$ & $4$ bits \\
\hline
$m=13$, $x=2$ & $0.5333$ & $8$ bits \\
\hline
$m=24$, $x=2$ & $0.5385$ & $14$ bits \\
\hline
$m=33$, $x=2$ & $0.5429$ & $19$ bits \\
\hline
$m=42$, $x=2$ & $0.5455$ & $24$ bits \\
\hline
$m=91$, $x=2$ & $0.5484$ & $51$ bits \\
\hline
\end{tabular}}
\label{table_4}
\vspace{-0.5em}
\end{table}

For the sake of comparison with other line codes having similar performance, we focus on constrained codes generated via finite-state machines (FSMs) and decoded via sliding window decoders \cite{siegel_mr, siegel_const, immink_1, ach_fsm} because of their practicality. The FSM-based constrained codes we compare with include both RLL and $\mathcal{T}_x$-constrained codes.

We briefly discuss $(d, k)$ RLL codes. An RLL code with parameter $d$ constrains each codeword to have at least $d$ $0$'s between each two consecutive $1$'s. RLL codes are used with NRZI signaling. Thus, an RLL code with parameter $d$ has any two consecutive transitions separated by at least $d+1$ successive bit durations\footnote{The maximum number of successive bit durations between two consecutive transitions for a $(d, k)$ RLL code with NRZI signaling is $k+1$. This maximum number is $k_{\textup{eff}}^{\textup{c}}$ for a C-LOCO code with NRZ signaling.}. Therefore, and from the definition of a LOCO code, an RLL code with parameter $d$ has similar performance to a LOCO code with parameter $x$.

Consider FSM-based RLL codes with $d=x$ and FSM-based $\mathcal{T}_x$-constrained codes. There are three main advantages of LOCO codes over FSM-based constrained codes used for the same purpose, which are:
\begin{enumerate}
\item LOCO codes achieve higher rates.
\item LOCO codes are immune against error propagation from a codeword into another.
\item Balancing LOCO codes is not only simple, but also incurs a very limited rate loss.
\end{enumerate}

The second and third advantages will be discussed later in this paper. As for the rate advantage, a practical FSM-based RLL code with $d=1$ typically has a rate of $0.6667$, which is the same rate a practical FSM-based $\mathcal{T}_1$-constrained code has \cite{siegel_mr, siegel_const}. This rate is lower than the rates of all C-LOCO codes with $x=1$ in Table~\ref{table_4} except the code with $m=8$. Moreover, a practical FSM-based RLL code with $d=2$ typically has a rate of $0.5000$, which is the same rate a practical FSM-based $\mathcal{T}_2$-constrained code has \cite{siegel_mr, immink_1}. This rate is lower than the rates of all C-LOCO codes with $x=2$ in Table~\ref{table_4} except the code with $m=6$. The rate gain of moderate-length C-LOCO codes over practical FSM-based constrained codes, where $d=x$, is up to $10\%$. In particular, $\mathcal{C}_{91,2}^{\textup{c}}$ achieves a rate of $0.5484$ at a moderate complexity, which is about $10\%$ higher than the typical rate of a practical FSM-based constrained code, where $d=x=2$, that is $0.5000$ (see also \cite{siegel_mr} and \cite{immink_1}).

The observation that constrained codes based on lexicographic indexing offer significant rate gains compared with FSM-based constrained codes was presented in \cite{immink_lex} and \cite{braun_lex}. However, the techniques proposed in both papers require the code length to be significantly large ($m > 250$) in order to achieve such gains, which is not needed for LOCO codes. This observation will be demonstrated even more upon introducing balanced LOCO codes.

We introduce now the encoding and decoding algorithms of our C-LOCO codes, which are based on Theorem~\ref{thm_induct}. Algorithm~\ref{alg_enc} is the encoding algorithm, and Algorithm~\ref{alg_dec} is the decoding algorithm.

\begin{algorithm}[H]
\caption{Encoding C-LOCO Codes}
\begin{algorithmic}[1]
\State \textbf{Input:} Incoming stream of binary messages.
\State Decide the value of $x$ and the bridging method based on system requirements.
\State Use (\ref{eqn_Ndef}) and (\ref{eqn_rec}) to compute $N(i, x)$, $i \in \{2, 3, \dots\}$.
\State Specify $m$, the smallest $i$ in Step~3 to achieve the desired rate. The message length is $s^{\textup{c}} = \left \lfloor \log_2 \left( N(m, x) - 2 \right )  \right \rfloor$.
\State \textbf{for} each incoming message $\bold{b}$ of length $s^{\textup{c}}$ \textbf{do}
\State \hspace{2ex} Compute $g(\bold{c})=\textup{decimal}(\bold{b})+1$. \textit{(binary sequence to decimal integer)}
\State \hspace{2ex} Initialize $\textup{residual}$ with $g(\bold{c})$.
\State \hspace{2ex} \textbf{if} $\textup{residual} < \frac{1}{2} N(m, x)$ \textbf{then}
\State \hspace{4ex} Encode $c_{m-1} = 0$.
\State \hspace{2ex} \textbf{else}
\State \hspace{4ex} Encode $c_{m-1} = 1$.
\State \hspace{4ex} $\textup{residual} \leftarrow \textup{residual} - \frac{1}{2} N(m, x)$.
\State \hspace{2ex} \textbf{end if}
\State \hspace{2ex} \textbf{for} $i \in \{m-2, m-3, \dots, 0\}$ \textbf{do} \textit{(in order)}
\State \hspace{4ex} \textbf{if} $\textup{residual} < \frac{1}{2} N(i-x+1, x)$ \textbf{then}
\State \hspace{6ex} Encode $c_i = 0$.
\State \hspace{4ex} \textbf{else}
\State \hspace{6ex} Initialize $\bold{f}$, which is a vector of $x$ entries, with $\bold{0}$. \textit{(the forbidden patterns indicators)}
\State \hspace{6ex} \textbf{if} $c_{i+1} = 0$ \textbf{then}
\State \hspace{8ex} $\beta_0 = \frac{1}{2} N(i-x+1, x)$.
\State \hspace{8ex} \textbf{for} $j \in \{1, 2, \dots, x\}$
\State \hspace{10ex} \textbf{if} $i-j < 0$ \textbf{then} \textit{(no forbidden patterns)}
\State \hspace{12ex} \textbf{break}. \textit{(exit the current loop)}
\State \hspace{10ex} \textbf{end if}
\State \hspace{10ex} $\beta_j = \beta_{j-1} + \frac{1}{2} N(i-x+1-j, x)$.
\State \hspace{10ex} \textbf{if} $\beta_{j-1} \leq \textup{residual} < \beta_j$ \textbf{then}
\State \hspace{12ex} $f_j = 1$. \textit{(a forbidden pattern of the form $0\bold{1}^j 0$ is spotted and has to be avoided)}
\State \hspace{12ex} \textbf{break}. \textit{(exit the current loop)}
\State \hspace{10ex} \textbf{end if}
\State \hspace{8ex} \textbf{end for}
\State \hspace{6ex} \textbf{end if}
\State \hspace{6ex} \textbf{if} $\bold{f} = \bold{0}$ \textbf{then} \textit{(no forbidden patterns)}
\State \hspace{8ex} Encode $c_i = 1$.
\State \hspace{8ex} $\textup{residual} \leftarrow \textup{residual} - \frac{1}{2} N(i-x+1, x)$.
\State \hspace{6ex} \textbf{else}
\State \hspace{8ex} Encode $c_i = 0$.
\State \hspace{6ex} \textbf{end if}
\State \hspace{4ex} \textbf{end if}
\State \hspace{2ex} \textbf{end for}
\State \hspace{2ex} Add $x$ bridging bits/symbols according to the bridging method. \textit{(the $x+1$ LMBs from the next codeword are needed here if the second bridging method is adopted)}
\State \textbf{end for}
\State \textbf{Output:} Outgoing stream of binary C-LOCO codewords.
\end{algorithmic}
\label{alg_enc}
\end{algorithm}

Consider the C-LOCO code $\mathcal{C}_{m,x}^{\textup{c}}$. It is possible that there exists a binary vector of length $m$, $\bold{e} \triangleq \left [ e_{m-1} \textup{ } e_{m-2} \textup{ } \dots \textup{ } e_0 \right ]$, which is not a C-LOCO codeword, and a C-LOCO codeword of length $m$, $\bold{c} \triangleq \left [ c_{m-1} \textup{ } c_{m-2} \textup{ } \dots \textup{ } c_0 \right ]$, such that:
\begin{align}
g(\bold{c}) &= \frac{1}{2} \left [ a_{m-1} N(m, x) + \sum_{i=0}^{m-2} a_i N(i-x+1, x) \right ] \allowdisplaybreaks \nonumber \\ &= \frac{1}{2} \left [ a^{\textup{e}}_{m-1} N(m, x) + \sum_{i=0}^{m-2} a^{\textup{e}}_i N(i-x+1, x) \right ],
\end{align}
where $a^{\textup{e}}_i$ is defined for each $e_i$ the same way $a_i$ is defined for each $c_i$ in (\ref{eqn_ai}). To prevent encoding a vector like $\bold{e}$, which is not a C-LOCO codeword, we need to prevent forbidden patterns from appearing while encoding via Algorithm~\ref{alg_enc}.

The steps from 18 to 31 in Algorithm~\ref{alg_enc} are to make sure forbidden patterns of the form $0\bold{1}^j 0$, $1 \leq j \leq x$, in $\mathcal{T}_x$ do not appear in any codeword. As for forbidden patterns of the form $1\bold{0}^j 1$, $1 \leq j \leq x$, they will never appear if forbidden patterns of the form $0\bold{1}^j 0$, $1 \leq j \leq x$, are guaranteed to be eliminated. The justification goes as follows. Suppose we are encoding $c_i$, $2x \leq i \leq m-2$, and $c_{i+1} = 1$. Since patterns of the form $0\bold{1}^j 0$, $1 \leq j \leq x$, do not appear in any codeword, it suffices to show that $1\bold{0}^j \bold{1}^{x+1}$, $1 \leq j \leq x$, cannot appear either. In other words, we want to show that if the variable $\textup{residual}$ is not enough to encode $c_i = 1$, it is not enough to encode $\left [ c_{i-1} \textup{ } c_{i-2} \textup{ } \dots \textup{ } c_{i-2x} \right ] = \bold{0}^{x-1} \bold{1}^{x+1}$, which implies that it is not enough to encode $\left [ c_{i-1} \textup{ } c_{i-2} \textup{ } \dots \textup{ } c_{i-x-j} \right ] = \bold{0}^{j-1} \bold{1}^{x+1}$, $1 \leq j \leq x$. This property for $\textup{residual}$ is satisfied if:
\begin{equation}\label{eqn_condalg}
\frac{1}{2} N(i-x+1, x) \leq \frac{1}{2} \sum_{\eta=x}^{2x} N(i-x+1-\eta, x).
\end{equation}

Let $\sigma \triangleq i-x+1$. From (\ref{eqn_rec}), we get:
\begin{align}\label{eqn_conpr}
&N(\sigma, x) = N(\sigma-1, x) + N(\sigma-x-1, x) \nonumber \\ &= N(\sigma-2, x) + N(\sigma-x-1, x) + N(\sigma-x-2, x) = \dots \nonumber \\ &= N(\sigma-x, x) + N(\sigma-x-1, x) + \dots + N(\sigma-2x, x) \nonumber \\ &= \sum_{\eta=x}^{2x} N(\sigma-\eta, x).
\end{align}
From (\ref{eqn_conpr}), we conclude that (\ref{eqn_condalg}) is true, and it is an equality, which means the $\textup{residual}$ property is satisfied. Note also that the conclusion is correct for $1 \leq i \leq 2x-1$, which completes the justification.

\begin{example}\label{ex_5}
We illustrate Algorithm~\ref{alg_enc} by showing how to encode a message using the C-LOCO code $\mathcal{C}_{6,1}^{\textup{c}}$ given in Table~\ref{table_3}. Here, $N(0, 1) \triangleq 2$, $N(1, 1) \triangleq 2$, $N(2, 1) = 4$, $N(3, 1) = 6$, $N(4, 1) = 10$, $N(5, 1) = 16$, and $N(6, 1) = 26$. Moreover, $s^{\textup{c}} = \left \lfloor \log_2 24  \right \rfloor = 4$. Consider the message $1110$. From Step~6, $g(\bold{c}) = \textup{decimal}(1110)+1 = 15$, which is the initial value of the variable $\textup{residual}$. Since $\textup{residual} > \frac{1}{2}N(6, 1) = 13$, from Step~11, $c_{5}$ is encoded as $1$. At Step~12, $\textup{residual}$ becomes $15 - 13 = 2$. Then, the algorithm enters the \textbf{for} loop from Step~14 to Step~39. The remaining $5$ bits of the codeword are encoded as follows:
\begin{itemize}
\item At $i=4$, $\textup{residual} < \frac{1}{2}N(4, 1) = 5$. Consequently, $c_{4}$ is encoded as $0$ at Step~16.
\item At $i=3$, $\textup{residual} < \frac{1}{2}N(3, 1) = 3$. Consequently, $c_{3}$ is encoded as $0$ at Step~16.
\item At $i=2$, $\textup{residual} = \frac{1}{2}N(2, 1) = 2$. Here, $c_3 = 0$. From Steps~20 and 25, $\beta_0 = \frac{1}{2}N(2, 1) = 2$ and $\beta_1 = \frac{1}{2}N(2, 1) + \frac{1}{2}N(1,1) = 3$, respectively. Since $\beta_0 = \textup{residual} < \beta_1$, the condition in Step~26 is satisfied, leading to $f_1 =1$, which means that if $c_2$ is encoded as $1$, a forbidden pattern of the form $010$ will be created on $c_3$, $c_2$, and $c_1$. Consequently, $c_{2}$ is encoded as $0$ at Step~36 to prevent this scenario.
\item At $i=1$, $\textup{residual} > \frac{1}{2}N(1, 1) = 1$. Here, $c_2 = 0$. From Steps~20 and 25, $\beta_0 = \frac{1}{2}N(1, 1) = 1$ and $\beta_1 = \frac{1}{2}N(1, 1) + \frac{1}{2}N(0,1) = 2$, respectively. Since $\beta_0 < \textup{residual} = \beta_1$, the condition in Step~26 is not satisfied, leading to $f_1 =0$. Consequently, $c_{1}$ is encoded as $1$ at Step~33, and $\textup{residual}$ becomes $2 - 1 = 1$.
\item At $i=0$, $\textup{residual} = \frac{1}{2}N(0, 1) = 1$. Here, $c_1 = 1$. Consequently, $c_{0}$ is encoded as $1$ at Step~33.
\end{itemize}
As a result, the codeword generated is $100011$, which is codeword indexed by $g(\bold{c})=15$ in Table~\ref{table_3}.
\end{example}

\begin{algorithm}[H]
\caption{Decoding C-LOCO Codes}
\begin{algorithmic}[1]
\State \textbf{Inputs:} Incoming stream of binary C-LOCO codewords, in addition to $m$, $x$, and $s^{\textup{c}}$.
\State Use (\ref{eqn_Ndef}) and (\ref{eqn_rec}) to compute $N(i, x)$, $i \in \{2, 3, \dots, m\}$.
\State \textbf{for} each incoming codeword $\bold{c}$ of length $m$ \textbf{do}
\State \hspace{2ex} Initialize $g(\bold{c})$ with $0$.
\State \hspace{2ex} \textbf{if} $c_{m-1} = 1$ \textbf{then}
\State \hspace{4ex} $g(\bold{c}) \leftarrow g(\bold{c}) + \frac{1}{2} N(m, x)$.
\State \hspace{2ex} \textbf{end if}
\State \hspace{2ex} \textbf{for} $i \in \{m-2, m-3, \dots, 0\}$ \textbf{do} \textit{(in order)}
\State \hspace{4ex} \textbf{if} $c_i = 1$ \textbf{then}
\State \hspace{6ex} $g(\bold{c}) \leftarrow g(\bold{c}) + \frac{1}{2} N(i-x+1, x)$.
\State \hspace{4ex} \textbf{end if}
\State \hspace{2ex} \textbf{end for}
\State \hspace{2ex} Compute $\bold{b}=\textup{binary}(g(\bold{c})-1)$, which has length $s^{\textup{c}}$. \textit{(decimal integer to binary sequence)}
\State \hspace{2ex} Ignore the next $x$ bridging bits/symbols.
\State \textbf{end for}
\State \textbf{Output:} Outgoing stream of binary messages.
\end{algorithmic}
\label{alg_dec}
\end{algorithm}

Example~\ref{ex_3} in Section~\ref{sec_pract} already showed how the decoding algorithm works.

As demonstrated by Algorithm~\ref{alg_enc} and Algorithm~\ref{alg_dec} in addition to Theorem~\ref{thm_induct}, the encoding procedure of C-LOCO codes is mainly comparisons and subtractions, while the decoding procedure of C-LOCO codes is mainly additions. The size of the adders used to perform these tasks, referred to in Tables~\ref{table_4} and \ref{table_7} as ``Adder size'', is $\log_2$ the maximum value $g(\bold{c})$ can take that corresponds to a message, and it is given by:
\begin{equation}
s^{\textup{c}} = \left \lfloor \log_2 \left( N(m, x) - 2 \right )  \right \rfloor,
\end{equation}
which is the message length. Table~\ref{table_4} links the rate of a C-LOCO code with its encoding and decoding complexity through the size of the adders to be used. For example, for a C-LOCO code with $x=1$, if a rate of $0.6667$ is satisfactory, small adders of size just $6$ bits are all what is needed. However, in case the rate needs to be about $0.6842$, adders of size $13$ bits should be used. Moreover, for a C-LOCO code with $x=2$, if a rate of $0.5000$ is satisfactory, small adders of size just $4$ bits are all what is needed. However, in case the rate needs to be about $0.5333$, adders of size $8$ bits should be used. Note that the cardinalities $N(i, x)$, $-x+1 \leq i \leq m$, should be stored in the memory offline. Note also that the multiplication by $\frac{1}{2}$ is just a right shift by one unit in binary, and it can be done only once at the beginning of the encoding-decoding.

From Table~\ref{table_4}, the C-LOCO code $\mathcal{C}_{90,1}^{\textup{c}}$ has rate $0.6923$ and adder size $63$ bits. The same rate is achieved in \cite{immink_table} for an RLL code with $d=1$ at code (resp., message) length $13$ (resp., $9$) bits. However, the technique in \cite{immink_table} is based on lookup tables; thus, the complexity of the encoding and decoding is governed by lookup tables of size $2^9 \times 13=6656$ bits. Note that in the case of $d=2$, the size of these lookup tables governing the complexity can reach $40960$ bits. This complexity is significantly higher than what we offer.

LOCO codes are also reconfigurable. In particular, if the size of the adders is appropriate, the same set of adders used to encode-decode a specific LOCO code can be reconfigured to encode-decode another LOCO code just by changing their inputs (the cardinalities) through multiplexers.

\begin{remark}\label{rmk_lorll_2}
Observe that (\ref{eqn_lorll_loco}) in Remark~\ref{rmk_lorll} shows that the capacity of a $\mathcal{T}_x$-constrained code is the same as the capacity of a $(d, \infty)$ RLL code with $d=x$ since:
\begin{equation}\label{eqn_cap}
\lim_{m \rightarrow \infty} \frac{\log_2 N(m, x)}{m} = \lim_{m \rightarrow \infty} \frac{\log_2 N_{\textup{RLL}}(m, d)}{m}.
\end{equation}
In other words, $(d, \infty)$ LO-RLL codes achieve similar rates to the rates of LOCO codes asymptotically. This fact can also be reached from the finite-state transition diagrams of the constraints \cite{siegel_const, immink_1}. However, (\ref{eqn_lorll_loco}) also shows that LOCO codes are more efficient compared with LO-RLL codes in the finite-length regime. The reason is that from (\ref{eqn_lorll_loco}) and (\ref{eqn_rec}), the difference between the cardinalities of a LOCO code $\mathcal{C}_{m,x}$ and a $(d, \infty)$ LO-RLL code with $d=x$ and length $m$ is:
\begin{align}
N(m, x) &- N_{\textup{RLL}}(m, d) = N(m, x) - \frac{1}{2} N(m+1, x) \nonumber \\ &= \frac{1}{2} \left [ N(m, x) - N(m-x, x) \right ].
\end{align}
Thus, if the same number of bits is used for bridging, the LOCO code can achieve higher rates at the same code length or lower complexities at the same rate\footnote{Another way to understand why this is the case is that for $d=x$ and at the same length, the $(d, \infty)$ RLL constraint results in forbidding more prospective codewords compared with the $\mathcal{T}_x$ constraint.}. This is also true when the two codes are self-clocked. For example, for $d=x=1$ and using $1$ bit/symbol for bridging in both codes, a self-clocked LOCO code of length $m=8$ and adder size of $6$ bits is enough to achieve a rate of $0.6667$ (see Table~\ref{table_4}), while to achieve the same rate using a self-clocked $(d, \infty)$ LO-RLL code, the length has to be increased to $m=17$ and the adder size to $12$ bits, which means roughly double the complexity of the LOCO encoding-decoding.
\end{remark}

We end this section by discussing two more aspects in the proposed LOCO codes: error propagation in addition to parallel encoding and decoding. The fixed length of LOCO codes makes them immune against error propagation from a codeword into the following ones. In particular, multiple errors occurring in one codeword do not affect the decoding of the following codewords. However, for large code lengths, few bit errors in a LOCO codeword can affect many bits in the message, which is the reason why we recommend LOCO codes with moderate lengths. On the contrary, FSM-based constrained codes with sliding window decoders suffer from error propagation among different codewords that is exacerbated with long codeword lengths (and also with long streams of codewords) \cite{howe_prop}. Furthermore, because of their fixed length, LOCO codes enable parallel encoding and decoding of different codewords if the complexity constraints of the system allow that. This advantage can be of significant value in data storage systems, where codewords are already written upon receiving (reading) them. On the other hand, FSM-based constrained codes with sliding window decoders do not enable efficient parallel encoding and decoding. The properties stated here for LOCO codes also apply to the balanced LOCO codes discussed in Section~\ref{sec_bala}.

%%%%%%%%%%%%%%%%%%%%%%%%%%%%%%%%%%%%%
\section{Density Gains in MR Systems}\label{sec_mr}

\begin{figure*}
\vspace{-0.2em}
\center
\includegraphics[trim={0.1in 1.0in 0.0in 1.2in}, width=7.3in]{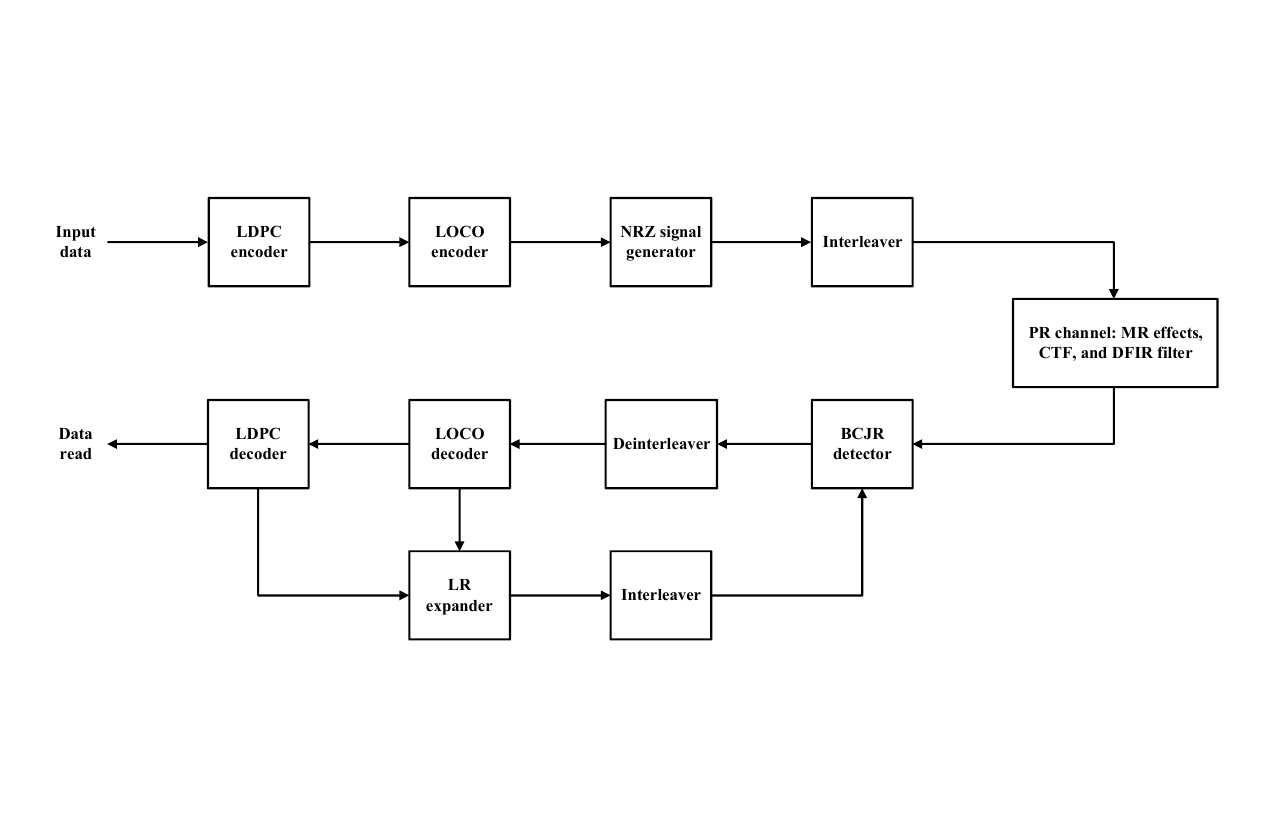}
\vspace{-1.6em}
\caption{MR system model with LDPC and LOCO codes used.}
\label{fig_1}
\vspace{-0.4em}
\end{figure*}

Our MR system model is shown in Fig.~\ref{fig_1}, and it consists of the following modules.

\textbf{LDPC encoder:} This is a binary spatially-coupled (SC) LDPC encoder, which takes $w$ bits of input data and generates an SC codeword of length $n$ bits. The adopted SC codes will be described shortly.

\textbf{LOCO encoder:} It takes the SC codeword as input, and using Algorithm~\ref{alg_enc}, it encodes only $n-w$ parity bits via a C-LOCO code $\mathcal{C}^{\textup{c}}_{m,x}$ to significantly increase their reliability by mitigating ISI for them as previously illustrated. The parameters of the C-LOCO code will be described shortly, but it has a much smaller length compared with $n-w$. Thus, there is a stream of C-LOCO codewords, with each consecutive two of them separated by a bridging pattern $\bold{z}^x$. The output of the LOCO encoder is of length $n_{\textup{ov}}$.

\textbf{NRZ signal generator:} It generates an NRZ stream of $n_{\textup{ov}}$ symbols, each of which is in $\{-A, +A\}$, except for the bridging symbols. Symbol $z$ for bridging corresponds to no transmission (no writing).

\textbf{Interleaver:} A pseudo-random interleaver is applied only on the $w$ bits that are not encoded via the C-LOCO code.

\textbf{PR channel:} We use the PR channel described in \cite{ahh_bas}. The MR channel effects are inter-symbol interference (intrinsic memory), jitter, and electronic noise. The channel density \cite{ahh_bas, shafa_2d}, which is the ratio of the read-head pulse duration at half the amplitudes to the bit duration, is swept to generate the plots. The signal-to-noise ratio (SNR) is $13.00$ dB. A continuous-time filter (CTF) followed by a digital finite-impulse-response (DFIR) filter are applied to achieve the PR equalization target $[8$~$14$~$2]$. Observe that this PR target, which is recommended by the industry, behaves in a way similar to the channel impulse response \cite{ahh_bas, shafa_2d}. This observation is an important reason why we are here adopting the set $\mathcal{T}_x$ of symmetric forbidden patterns, which is closed under taking pattern complements.

\textbf{BCJR detector:} A Bahl Cocke Jelinek Raviv (BCJR) detector \cite{bcjr}, which is based on pattern-dependent noise prediction (PDNP) \cite{pdnp}, is then applied to the received stream to calculate $n_{\textup{ov}}$ likelihood ratios (LRs). There is a feedback loop incorporating the detector and the decoders.

\textbf{Deinterleaver:} It rearranges the LRs of the $w$ bits that were not encoded via the C-LOCO code, i.e., the ones that were originally interleaved.

\textbf{LOCO decoder:} Initially, this decoder makes a hard decision on the $n_{\textup{ov}} - w$ bits that were encoded via the C-LOCO code using their LRs. If the $\mathcal{T}_x$ constraint is violated for the received word or the received word is in $\{\bold{0}^m, \bold{1}^m\}$, the LOCO decoder tries to fix that by flipping the bit with the closest LR to $1$ (the smallest $\log_e$ LR in magnitude). In other words, the LOCO decoder performs some sort of error correction here. Next, it decodes the original $n - w$ parity bits using Algorithm~\ref{alg_dec}. Finally, the LOCO decoder sends $n$ LRs to the LDPC decoder; $w$ LRs left as they are, and $n-w$ highly reliable LRs.

\textbf{LDPC decoder:} This is a fast Fourier transform based $q$-ary~sum-product algorithm (FFT-QSPA) LDPC decoder \cite{dec_fft}, with $q$, the GF order, being set to $2$ here. The number of global (detector-decoders) iterations is $10$, and the number of local (LDPC decoder only) iterations is $20$. Unless a codeword is reached, the LDPC decoder performs its prescribed number of local iterations for each global iteration. At the end of each global iteration, except the last one, the LDPC decoder, sends its updated $n$ LRs in the feedback loop.

\textbf{LR expander:} The BCJR detector operates on $n_{\textup{ov}}$ symbols. Thus, an LR expander is used to expand the LR vector from $n$ to $n_{\textup{ov}}$ via the information it receives from the LOCO and the LDPC decoders.

\textbf{Interleaver:} The interleaver in the feedback branch of the detector-decoders loop is a pseudo-random interleaver, which is applied only on the $w$ LRs of the bits that were not encoded via the C-LOCO code.

At the last global iteration, looping stops, and the LDPC decoder generates the data read ($w$ bits). More details about some of these modules can be found in \cite{ahh_bas}.

\begin{remark}
If the C-LOCO message length, $s^{\textup{c}}$, does not divide $n-w$, we pad with few, say $\delta$, zeros.
\end{remark}

One of the two reasons why we do not apply the C-LOCO code on the entire LDPC codeword here is to limit the rate loss resulting from integrating the C-LOCO code in the MR system, which is a critical requirement in all data storage systems. The other reason will be introduced upon discussing the simulation plots. Lemma~\ref{lem_rate_mr} gives the overall rate of the LDPC-LOCO coding scheme applied in our system.

\begin{lemma}\label{lem_rate_mr}
Consider the following LDPC-LOCO coding scheme. A C-LOCO code of rate $R_{\textup{LOCO}}^{\textup{c}}$ is used to encode only the parity bits of an LDPC code of rate $R_{\textup{LDPC}}$. The overall rate of this scheme is:
\begin{equation}\label{eqn_rate_ov}
R_{\textup{ov}} \approx \frac{R_{\textup{LDPC}} R_{\textup{LOCO}}^{\textup{c}}}{R_{\textup{LDPC}} R_{\textup{LOCO}}^{\textup{c}} + (1 - R_{\textup{LDPC}})}.
\end{equation}
\end{lemma}

\begin{IEEEproof}
The length of the LDPC codeword can be written as:
\begin{equation}
n = w + (n-w).
\end{equation}
Only those $n-w$ bits are going to be encoded via the C-LOCO code. Consequently,
\begin{equation}\label{eqn_ovlen}
n_{\textup{ov}} = w + (n-w+\delta)\frac{1}{R_{\textup{LOCO}}^{\textup{c}}}.
\end{equation}
As a result, the overall rate is:
\begin{align}\label{eqn_ovrate_pr}
R_{\textup{ov}} &= \frac{w}{n_{\textup{ov}}} = \frac{w}{w + (n-w+\delta)\frac{1}{R_{\textup{LOCO}}^{\textup{c}}}} \nonumber \\ &= \frac{w/n}{w/n + (1-w/n+\delta/n)\frac{1}{R_{\textup{LOCO}}^{\textup{c}}}} \nonumber \\ &\approx \frac{R_{\textup{LDPC}}}{R_{\textup{LDPC}} + (1-R_{\textup{LDPC}})\frac{1}{R_{\textup{LOCO}}^{\textup{c}}}} \nonumber \\ &= \frac{R_{\textup{LDPC}} R_{\textup{LOCO}}^{\textup{c}}}{R_{\textup{LDPC}} R_{\textup{LOCO}}^{\textup{c}} + (1 - R_{\textup{LDPC}})}.
\end{align}
Note that $\delta$ is very small compared with $n$.
\end{IEEEproof}

Lemma~\ref{lem_rate_mr} demonstrates that the rate loss due to integrating a C-LOCO code in the MR system the way we do it is limited. In fact, from the expression in (\ref{eqn_rate_ov}), as $R_{\textup{LDPC}}$ approaches $1$, $R_{\textup{ov}}$ approaches $R_{\textup{LDPC}}$. The reason is that when $R_{\textup{LDPC}}$ approaches $1$, $R_{\textup{ov}}$ becomes $h/(h+\epsilon)$, where $\epsilon = 1 - R_{\textup{LDPC}} << h = R_{\textup{LDPC}}R^{\textup{c}}_{\textup{LOCO}}$. Thus, $R_{\textup{ov}}$ also approaches $1$ like $R_{\textup{LDPC}}$. Numerical examples are: for $R_{\textup{LDPC}} = 0.7000$ and $R^{\textup{c}}_{\textup{LOCO}}=0.6667$, $R_{\textup{ov}} = 0.6087$, while for $R_{\textup{LDPC}} = 0.9500$ and $R^{\textup{c}}_{\textup{LOCO}}=0.6667$, $R_{\textup{ov}} = 0.9268$, which is only $2.4\%$ lower than $R_{\textup{LDPC}}$.

\begin{figure*}
\vspace{-0.2em}
\center
\includegraphics[trim={1.5in 0.7in 1.0in 0.9in}, width=4.3in]{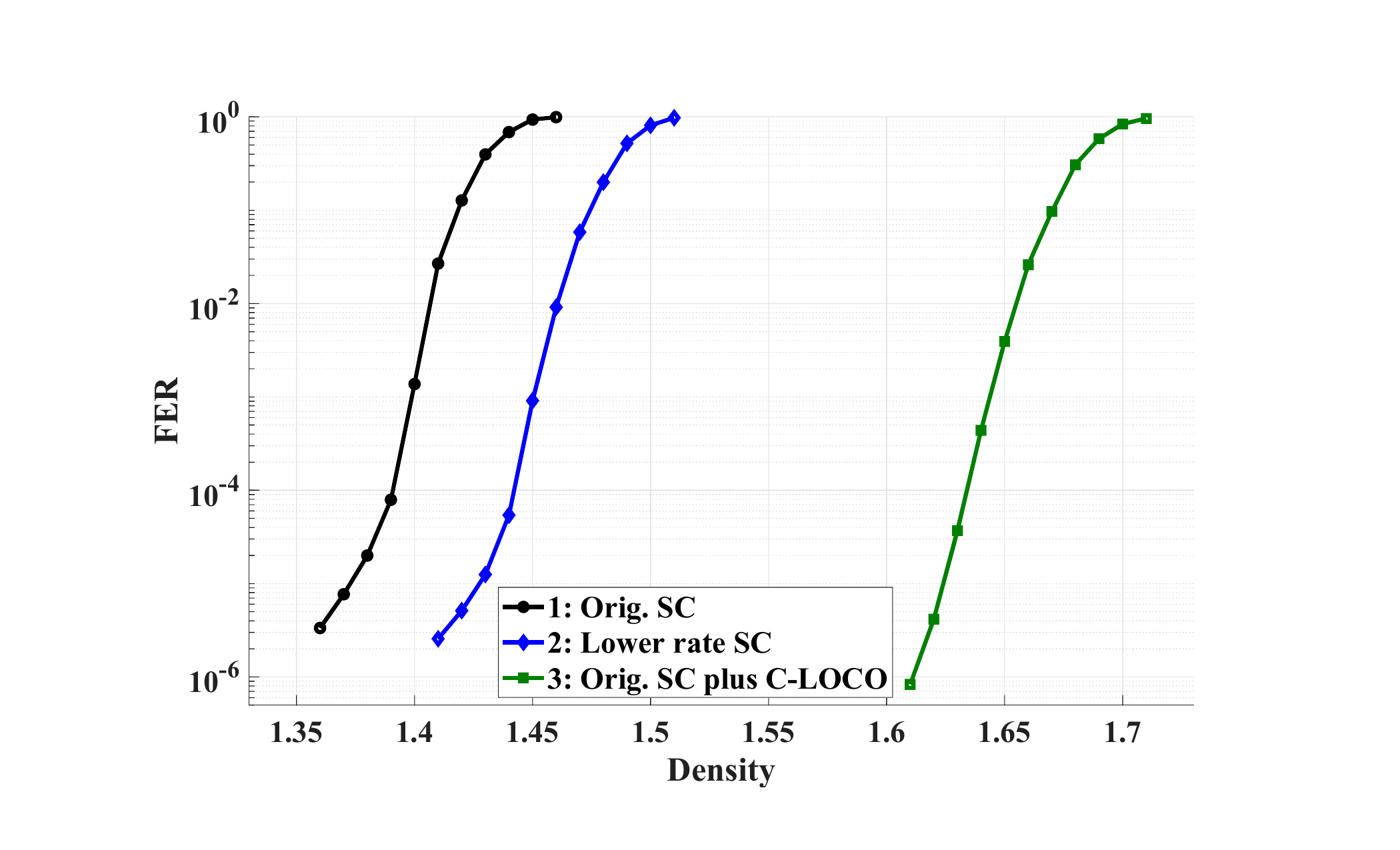}
\vspace{-0.0em}
\caption{Density gains achieved by LOCO codes in MR systems.}
\label{fig_2}
\vspace{-0.3em}
\end{figure*}

There are two binary SC codes used in our simulations. The two codes are constructed according to \cite{ahh_tit2}, which provides a method to design high performance SC codes particularly for MR systems. This method is based on the optimal overlap, circulant power optimizer (OO-CPO) approach. SC~Code~1 has column weight $=4$, maximum row weight $=17$, circulant size $=37$, memory $=1$, and coupling length $=6$. Thus, SC~Code~1 has block length $=3774$ bits and rate $\approx 0.725$. SC~Code~2 has column weight $=4$, maximum row weight $=13$, circulant size $=47$, memory $=1$, and coupling length $=7$. Thus, SC~Code~2 has block length $=4277$ bits and rate $\approx 0.648$. The differences in length and rate between the two SC codes will be illustrated shortly. Only SC~Code~1 will be combined with a C-LOCO code.

The C-LOCO code we use in the simulations is the code $\mathcal{C}_{18,1}^{\textup{c}}$. This code has $m = 18$ and $x = 1$. Thus, from (\ref{eqn_keff}), $\mathcal{C}_{18,1}^{\textup{c}}$ has $k_{\textup{eff}}^{\textup{c}} = 2 \times 17 + 1 = 35$.  Moreover, $\mathcal{C}_{18,1}^{\textup{c}}$ has $N^{\textup{c}}(18, 1) = 8362$, which means the message length is $s^{\textup{c}} = \left \lfloor \log_2 8360 \right \rfloor = 13$. Thus, from (\ref{eqn_rate}), the rate of $\mathcal{C}_{18,1}^{\textup{c}}$ is $\frac{13}{18+1} = 0.6842$ since one symbol $z$ is used for bridging.

We generate three plots, as shown in Fig.~\ref{fig_2}, for the following three simulation setups:
\begin{enumerate}
\item SC~Code~1 (original SC code) is used for error correction, and no C-LOCO code is applied.
\item SC~Code~2 (lower rate SC code) is used for error correction, and no C-LOCO code is applied.
\item SC~Code~1 is combined with the C-LOCO code $\mathcal{C}_{18,1}^{\textup{c}}$ such that only the parity bits of SC~Code~1 are encoded via $\mathcal{C}_{18,1}^{\textup{c}}$.
\end{enumerate}
The energy per input data bit in all three setups is the same.

For Setup~3, we have the following parameters: $w = 2738$ (see \cite{ahh_tit2}), $n = 3774$, $R_{\textup{LDPC}} = 0.725$, $R_{\textup{LOCO}}^{\textup{c}} = 0.6842$, and $\delta = 4$. From (\ref{eqn_ovlen}), the overall length after applying the C-LOCO code in Setup~3 is:
\begin{equation}
n_{\textup{ov}} = 2738 + (1036+4)\frac{1}{0.6842} = 4258. \nonumber
\end{equation}
Furthermore, from (\ref{eqn_rate_ov}), the overall rate is $R_{\textup{ov}} \approx 0.643$. Thus, the overall length and rate in Setup~3 are similar to the length and rate of SC~Code~2 in Setup~2.

The frame error rate (FER) versus density plots for the three setups are shown in Fig.~\ref{fig_2}. The figure demonstrates the gains of Setup~3, in which a C-LOCO code is applied in the MR system, over the other two setups. In particular, the density gain of Setup~3 over Setup~1 (resp., Setup~2) is about $20\%$ (resp., $16\%$) at FER $\approx 10^{-6}$. The density gain achieved in Setup~3 over Setup~2 implies that exploiting the additional redundancy by applying a C-LOCO code is significantly more helpful compared with exploiting this redundancy by adding more parity bits. An intriguing observation from Fig.~\ref{fig_2} is that the error floor slope in Setup~3 is sharper than the error floor slope in the other two setups.

While applying the C-LOCO code to the entire LDPC codeword provides higher density gains, the overall rate loss becomes very high since the rate in this case becomes $R_{\textup{ov}} \approx R_{\textup{LDPC}} R_{\textup{LOCO}}^{\textup{c}}$. For example, if $\mathcal{C}_{18,1}^{\textup{c}}$ is applied to the entire codeword of SC~Code~1, the overall rate becomes $R_{\textup{ov}} \approx 0.496$, which is a lot lower than $R_{\textup{ov}}$ in Setup~3, which is $0.643$. Additionally, the density gains achieved diminish gradually with more bits being encoded via the C-LOCO code. In summary, the proposed idea in Setup~3 offers a better rate-density gain trade-off.

Setup~3 is motivated by a particular understanding of graph-based codes. Even though only a group of bits in the LDPC codeword, which are the bits encoded via the LOCO code, have highly reliable LRs while decoding, the information in these highly reliable LRs will be spread to all bits during the message passing procedure. Therefore, the LDPC decoder experiences a version of the channel with a higher effective~SNR, which results in the decoder, aided by the detector and the LOCO decoder, kicking-off its operation at higher densities.

The contribution in Section~\ref{sec_mr} is the idea that investing the additional redundancy in protecting the parity bits only of an LDPC code from ISI is significantly more effective than investing this redundancy in adding more parity bits. Observe that if we apply the same setup but replace the LOCO code with an RLL code having $d=x$ and the same rate, the performance gains would be comparable. However, there will be an additional complexity associated with using an RLL code that has the same rate as the LOCO code, which is discussed in Sections~\ref{sec_ralg} and \ref{sec_bala} in the paper.

\begin{remark}
In this paper, we use the word ``moderate'' to describe lengths of LOCO codes. The context of this usage may not be generalized to include LDPC codes since what is moderate for LOCO codes is very small for LDPC codes.
\end{remark}

%%%%%%%%%%%%%%%%%%%%%%%%%%%%%%%%%%%%%
\section{Balanced LOCO Codes}\label{sec_bala}

A critical additional requirement in line codes, which appears in applications like optical recording, Flash memories, in addition to USB and PCIe standards, is balancing \cite{knuth_bal, qin_flash, braun_lex}. Examples of balanced line codes are the $8$b/$10$b \cite{immink_2} and the $64$b/$66$b \cite{walker_66} codes (the latter is not strictly DC-free). Balanced line codes have zero average power at frequency zero, i.e., no DC power component, when the signal levels are $-A$ and $+A$. This is achieved by constraining the running disparity $p_{\textup{r}}$ of any stream of codewords from the line code. The work in \cite{robert_spec1} relates the running disparity to the width of the power spectral null. The running disparity $p_{\textup{r}}$ is measured before each new codeword in the stream, and $p_{\textup{r}}$ equals the sum of disparities of all the previous codewords and their bridging patterns. The disparity of a codeword $\bold{c}$, $p(\bold{c})$, is defined as the difference between the number of $+A$ and $-A$ ($+A$ and $E$ in Flash) symbols in the transmitted (written) codeword after the signaling scheme is applied. When NRZ signaling is applied, this disparity is directly the difference between the number of $1$'s and $0$'s in the codeword.

A standard way of balancing line codes is to encode each message to one of two codewords having the same magnitude but opposite signs for their disparities. Then, depending on the sign of the running disparity, one of these two codewords is picked for the incoming message. Codewords having zero disparity can be used to uniquely encode messages. For example, the $8$b/$10$b code adopts this way of balancing. This simple code is constructed to achieve balancing and self-clocking only, which is why it has a high rate. More advanced line codes, e.g., $\mathcal{T}_x$-constrained or RLL codes, have more requirements, e.g., improving the performance in data storage systems, making their rates less compared with the above simple line code. Thus, balancing these constrained codes via the approach mentioned in this paragraph incurs a penalty. This penalty is either rate loss (rate reduction) for the same complexity or additional complexity for the same rate.

In this section, we demonstrate another advantage of LOCO codes, which is that they can be balanced with the minimum penalty. We start with the following lemma.

\begin{lemma}\label{lem_inv}
Define codeword $\bold{c}^0$ as a LOCO codeword in $\mathcal{C}_{m,x}$ that starts with $0$ from the left. Define codeword $\bold{c}^1$ as the LOCO codeword indexed by $N(m,x)-1-g(\bold{c}^0)$ in $\mathcal{C}_{m,x}$, where $g(\bold{c}^0)$ is the index of $\bold{c}^0$. The two codewords $\bold{c}^0$ and $\bold{c}^1$ are the complements of each other.
\end{lemma}

\begin{IEEEproof}
We first define $a^0_i$ (resp., $a^1_i$) for each bit $c^0_i$ in $\bold{c}^0$ (resp.,~$c^1_i$ in $\bold{c}^1$) as in (\ref{eqn_ai}).

Since $\bold{c}^0$ starts with $0$ from the left, using (\ref{eqn_g_all}) gives:
\begin{equation}\label{eqn_invp}
g(\bold{c}^0) = \frac{1}{2} \left [ 0 + \sum_{i=0}^{m-2} a^0_i N(i-x+1, x) \right ].
\end{equation}
From the definition of $\bold{c}^1$, it has to start with $1$ from the left. Thus, using (\ref{eqn_g_all}) gives:
\begin{equation}\label{eqn_invm}
g(\bold{c}^1) = \frac{1}{2} \left [ N(m, x) + \sum_{i=0}^{m-2} a^1_i N(i-x+1, x) \right ].
\end{equation}
Furthermore, we also have:
\begin{equation}
g(\bold{c}^1) \triangleq N(m,x)-1-g(\bold{c}^0).
\end{equation}
Consequently, using (\ref{eqn_invp}) and (\ref{eqn_invm}), we get:
\begin{align}
g(\bold{c}^1) &+ g(\bold{c}^0) = \frac{1}{2} \sum_{i=0}^{m-2} a^0_i N(i-x+1, x) \nonumber \\ &+ \frac{1}{2} \left [ N(m, x) + \sum_{i=0}^{m-2} a^1_i N(i-x+1, x) \right ] \nonumber \\ &\hspace{+3.5em} = N(m,x)-1,
\end{align}
which means:
\begin{align}\label{eqn_blast}
\frac{1}{2} \sum_{i=0}^{m-2} a^0_i N(i-x+1, x) &+ \frac{1}{2}\sum_{i=0}^{m-2} a^1_i N(i-x+1, x) \nonumber \\ &\hspace{-8.5em} = \frac{1}{2}N(m,x)-1 = \frac{1}{2} \sum_{i=0}^{m-2} N(i-x+1,x).
\end{align}
The last equality in (\ref{eqn_blast}) follows from that $\frac{1}{2}N(m,x)-1$ is the index of the LOCO codeword $0\bold{1}^{m-1}$.
 
For a given codeword $\bold{c}^0$ starting with $0$ from the left in $\mathcal{C}_{m,x}$, the codeword $\bold{c}^1$ starting with $1$ from the left in $\mathcal{C}_{m,x}$, and having the $m-1$ RMBs being the complements of the $m-1$ RMBs in $\bold{c}^0$, makes (\ref{eqn_blast}) satisfied. Because the mapping from $g(\bold{c}^1)$ to $\bold{c}^1$ is one-to-one, such a codeword has to be the only codeword with that property. Since $c^0_{m-1}=0$ and $c^1_{m-1}=1$ are already complements, $\bold{c}^0$ and $\bold{c}^1$ are then the complements of each other.
\end{IEEEproof}

Note that since we adopt NRZ signaling,
\begin{equation}\label{eqn_disp}
p(\bold{c}^0) = - p(\bold{c}^1).
\end{equation}
Thus, and based on Lemma~\ref{lem_inv}, we now define the proposed balanced LOCO (B-LOCO) codes.

\begin{definition}\label{def_bloco}
A balanced LOCO (B-LOCO) code $\mathcal{C}_{m, x}^{\textup{b}}$, with $m \geq 2$, is a LOCO code in which, each pair of codewords $\bold{c}^0$~and $\bold{c}^1$, having indices $g(\bold{c}^0)$ and $g(\bold{c}^1) \triangleq N(m,x)-1-g(\bold{c}^0)$ in $\mathcal{C}_{m, x}$, respectively, are used to encode a single message. The selected codeword $\bold{c}$ is either $\bold{c}^0$ or $\bold{c}^1$ depending on the sign of the running disparity $p_{\textup{r}}$ as shown in Table~\ref{table_5}. Consequently, the cardinality of $\mathcal{C}_{m, x}^{\textup{b}}$ is:
\begin{equation}\label{eqn_bcl_card}
N^{\textup{b}}(m, x) = N(m, x).
\end{equation}
However, only a maximum of $\frac{1}{2}N^{\textup{b}}(m, x)$ codewords in $\mathcal{C}_{m, x}^{\textup{b}}$ correspond to distinct messages\footnote{That is why the minimum length we adopt for a B-LOCO code, and later a self-clocked B-LOCO code, is the length at which the cardinality $= 4$.}.

\begin{table}
\caption{The selection criterion for balancing in a B-LOCO code $\mathcal{C}_{m,x}^{\textup{b}}$. If $p_{\textup{r}} = 0$ or/and $p(\bold{c}^0) = p(\bold{c}^1) = 0$, select $\bold{c} = \bold{c}^0$.}
\vspace{-0.5em}
\centering
\scalebox{1.00}
{
\begin{tabular}{|c|c|}
\hline
\makecell{$\textup{sign}(p_{\textup{r}})$} & \makecell{Selected codeword $\bold{c}$} \\
\hline
$+$ & $\bold{c}^0$ or $\bold{c}^1$ such that $\textup{sign}(p(\bold{c}))$ is $-$  \\
\hline
$-$ & $\bold{c}^0$ or $\bold{c}^1$ such that $\textup{sign}(p(\bold{c}))$ is $+$ \\
\hline
\end{tabular}}
\label{table_5}
\vspace{-0.5em}
\end{table}
\end{definition}

\begin{remark}
If the second bridging method is adopted and $p_{\textup{r}} = 0$ or/and $p(\bold{c}^0) = p(\bold{c}^1) = 0$, it is also possible to select the codeword that enhances self-clocking taking into account the previous codeword.
\end{remark}

\begin{example}\label{ex_6}
The B-LOCO code $\mathcal{C}_{6,1}^{\textup{b}}$ is shown in Table~\ref{table_6} with the codeword disparities. Observe that (\ref{eqn_disp}) is always satisfied, i.e., $p(\bold{c}^0) = - p(\bold{c}^1)$. The cardinality of $\mathcal{C}_{6,1}^{\textup{b}}$ is:
\begin{equation}\label{eqn_ex6}
N^{\textup{b}}(6, 1) = N(6, 1) = 26. \nonumber
\end{equation}
However, only a maximum of $13$ codewords in $\mathcal{C}_{6, 1}^{\textup{b}}$ correspond to distinct messages.

\begin{table}
\caption{The B-LOCO code $\mathcal{C}_{6,1}^{\textup{b}}$. The CB-LOCO code $\mathcal{C}_{6,1}^{\textup{cb}}$ for all messages is the rows having $g^{\textup{b}}(\bold{c}) \in \{1, 2, \dots, 8\}$.}
\vspace{-0.5em}
\centering
\scalebox{1.00}
{
\begin{tabular}{|c|c|c|c|c|c|}
\hline
\makecell{Message} & \makecell{$g^{\textup{b}}(\bold{c})$} & \makecell{$\bold{c}^0$} & \makecell{$p(\bold{c}^0)$} & \makecell{$\bold{c}^1$} & \makecell{$p(\bold{c}^1)$} \\
\hline
{ } & $0$ & $000000$ & $-6$ & $111111$ & $+6$ \\
\hline
$000$ & $1$ & $000001$ & $-4$ & $111110$ & $+4$ \\
\hline
$001$ & $2$ & $000011$ & $-2$ & $111100$ & $+2$ \\
\hline
$010$ & $3$ & $000110$ & $-2$ & $111001$ & $+2$ \\
\hline
$011$ & $4$ & $000111$ & $0$ & $111000$ & $0$ \\
\hline
$100$ & $5$ & $001100$ & $-2$ & $110011$ & $+2$ \\
\hline
$101$ & $6$ & $001110$ & $0$ & $110001$ & $0$ \\
\hline
$110$ & $7$ & $001111$ & $+2$ & $110000$ & $-2$ \\
\hline
$111$ & $8$ & $011000$ & $-2$ & $100111$ & $+2$ \\
\hline
{ } & $9$ & $011001$ & $0$ & $100110$ & $0$ \\
\cline{2-6}
{ } & $10$ & $011100$ & $0$ & $100011$ & $0$ \\
\cline{2-6}
{ } & $11$ & $011110$ & $+2$ & $100001$ & $-2$ \\
\cline{2-6}
{ } & $12$ & $011111$ & $+4$ & $100000$ & $-4$ \\
\hline
\end{tabular}}
\label{table_6}
\vspace{-0.3em}
\end{table}
\end{example}

The running disparity in the case of B-LOCO codes satisfies $-m \leq p_{\textup{r}} < +m$ (see also Example~\ref{ex_6}). In particular, $-m \leq p_{\textup{r}} \leq +m-2$ if $m$ is even, and $-m \leq p_{\textup{r}} \leq +m-1$ if $m$ is odd. Moreover, because of the way codewords are chosen, as shown in Table~\ref{table_5}, this running disparity is around $0$ most of the time for long streams of codewords.

The following theorem is the key theorem for encoding and decoding B-LOCO codes.

\begin{theorem}\label{thm_induct2}
Consider a B-LOCO code $\mathcal{C}_{m,x}^{\textup{b}}$ with $m \geq 2$.~The index $g^{\textup{b}}(\bold{c})$ of a codeword $\bold{c} \in \mathcal{C}_{m,x}^{\textup{b}}$ is derived from $\bold{c}$ itself according to the following two equations:

\noindent If the LMB $c_{m-1} = 0$:
\begin{equation}\label{eqn_g20}
g^{\textup{b}}(\bold{c}) = \frac{1}{2} \sum_{i=0}^{m-2} a_i N(i-x+1, x).
\end{equation}
If the LMB $c_{m-1} = 1$:
\begin{equation}\label{eqn_g21}
g^{\textup{b}}(\bold{c}) = \frac{1}{2} \sum_{i=0}^{m-2} (1-a_i) N(i-x+1, x).
\end{equation}
Here, we use the abbreviated notation $g^{\textup{b}}(\bold{c})$ for simplicity.
\end{theorem}

\begin{IEEEproof}
For the case of $c_{m-1} = 0$, it is clear that:
\begin{equation}
g^{\textup{b}}(\bold{c}) = g(\bold{c}^0),
\end{equation}
where $g(\bold{c}^0)$ is the index of $\bold{c}^0$ in $\mathcal{C}_{m,x}$. Thus, using (\ref{eqn_g_all}):
\begin{align}\label{eqn_prf1}
g^{\textup{b}}(\bold{c}) &= \frac{1}{2} \left [ 0 + \sum_{i=0}^{m-2} a^0_i N(i-x+1, x) \right ] \nonumber \\ &= \frac{1}{2} \sum_{i=0}^{m-2} a_i N(i-x+1, x).
\end{align}

For the case of $c_{m-1} = 1$, $g^{\textup{b}}(\bold{c})$ must equal that of the corresponding codeword in $\mathcal{C}_{m,x}^{\textup{b}}$ that starts with $0$ from the left. From Lemma~\ref{lem_inv}, $\bold{c}$ in $\mathcal{C}_{m,x}^{\textup{b}}$ that has $c_{m-1} = 1$, which is  $\bold{c}^1$ in $\mathcal{C}_{m,x}$, and its corresponding codeword in $\mathcal{C}_{m,x}^{\textup{b}}$ that starts with $0$ from the left, which is  $\bold{c}^0$ in $\mathcal{C}_{m,x}$, are the complements of each other. Consequently, we conclude:
\begin{align}\label{eqn_prf2}
g^{\textup{b}}(\bold{c}) &= \frac{1}{2} \sum_{i=0}^{m-2} a^0_i N(i-x+1, x) \nonumber \\ &= \frac{1}{2} \sum_{i=0}^{m-2} (1-a^1_i) N(i-x+1, x) \nonumber \\ &= \frac{1}{2} \sum_{i=0}^{m-2} (1-a_i) N(i-x+1, x),
\end{align}
which completes the proof.
\end{IEEEproof}

\begin{example}\label{ex_7}
We illustrate Theorem~\ref{thm_induct2} via an example. Consider $\mathcal{C}_{6,1}^{\textup{b}}$ given in Table~\ref{table_6}. We check the two codewords indexed by $6$, which are $001110$ and $110001$. From (\ref{eqn_g20}), the codeword starting with $0$ from the left has:
\begin{align}
g^{\textup{b}}(\bold{c}) &= \frac{1}{2} \sum_{i=0}^{4} a_i N(i, 1) \nonumber \\ &= \frac{1}{2} \left [ N(3, 1) + N(2, 1) + N(1, 1) \right ] \nonumber \\ &= \frac{1}{2} \left [ 6 + 4 + 2 \right ] = 6. \nonumber
\end{align}
From (\ref{eqn_g21}), the codeword starting with $1$ from the left has:
\begin{align}
g^{\textup{b}}(\bold{c}) &= \frac{1}{2} \sum_{i=0}^{4} (1-a_i) N(i, 1) \nonumber \\ &= \frac{1}{2} \left [ N(3, 1) + N(2, 1) + N(1, 1) \right ] \nonumber \\ &= \frac{1}{2} \left [ 6 + 4 + 2 \right ] = 6. \nonumber
\end{align}
\end{example}

Bridging in B-LOCO codes is performed the same way as described in Section~\ref{sec_ralg} for LOCO codes. Define the disparity change resulting from adding a $z$ symbol after a B-LOCO codeword to be $0$, which makes sense as $z$ is the no transmission (no writing) symbol. Observe that whether the first method or the second method is used for bridging, the above analysis does not change. This statement is clear for the first method. As for the second method, note that the complement rule in Lemma~\ref{lem_inv} applies also for bridging patterns (see Table~\ref{table_2}), which justifies the statement. We use the first bridging method in this section since, in addition to its simplicity, it results in no disparity change, and thus no increase in the maximum magnitude of the running disparity.

\begin{definition}\label{def_clob}
A self-clocked B-LOCO (CB-LOCO) code $\mathcal{C}_{m,x}^{\textup{cb}}$ is the code resulting from removing the all $0$'s and the all $1$'s codewords from the B-LOCO code $\mathcal{C}_{m,x}^{\textup{b}}$. In particular,
\begin{equation}\label{eqn_sclb}
\mathcal{C}_{m,x}^{\textup{cb}} \triangleq \mathcal{C}_{m,x}^{\textup{b}} \setminus \{\bold{0}^m,  \bold{1}^m\},
\end{equation}
where $m \geq 3$. The cardinality of $\mathcal{C}_{m,x}^{\textup{cb}}$ is given by:
\begin{equation}\label{eqn_sclb_card}
N^{\textup{cb}}(m, x) = N^{\textup{b}}(m, x) - 2 = N(m, x) - 2.
\end{equation}
However, only a maximum of $\frac{1}{2}N^{\textup{cb}}(m, x)$ codewords in $\mathcal{C}_{m, x}^{\textup{cb}}$ correspond to distinct messages.
\end{definition}

Define $k_{\textup{eff}}^{\textup{cb}}$ as the maximum number of successive bit durations between two consecutive transitions in a stream of CB-LOCO codewords that belong to $\mathcal{C}_{m,x}^{\textup{cb}}$, with each two consecutive codewords separated by $\bold{z}^x$. Recall that a transition is only from $0$ to $1$ or from $1$ to $0$. Consequently, we get:
\begin{equation}\label{eqn_kbeff}
k_{\textup{eff}}^{\textup{cb}} = k_{\textup{eff}}^{\textup{c}} = 2(m-1) + x.
\end{equation}

\begin{remark}
A stream of B-LOCO codewords that belong to $\mathcal{C}_{m,x}^{\textup{b}}$, each having $g^{\textup{b}}(\bold{c})=0$ and using the first bridging method, is encoded as follows:
\begin{align}
\bold{0}^{m} - \bold{z}^x - \bold{1}^{m} - \bold{z}^x - \bold{0}^{m} -  \bold{z}^x -  \bold{1}^m - \dots. \nonumber
\end{align}
If the system can make use of the $0 - z$ (resp., $z - 1$) followed by the $z - 1$ (resp., $0 - z$) changes for self-clocking, the two codewords $\bold{0}^m$ and $\bold{1}^m$ can be kept in the code. Here, we assume that the system cannot use these changes for self-clocking, and that is why our definition for a transition is exclusively from $0$ to $1$ or from $1$ to $0$.
\end{remark}

Note that the maximum magnitude of the running disparity in the case of CB-LOCO codes is $m-2$, not $m$, because of the removal of the two codewords $\bold{0}^{m}$ and $\bold{1}^{m}$. Thus, CB-LOCO codes are better than B-LOCO codes in that regard.

\begin{remark}
If the second bridging method is used instead, the two codewords $\bold{0}^m$ and $\bold{1}^m$ can be kept in the code, and $k_{\textup{eff}}^{\textup{b}}$ becomes $\lfloor 5(m+x)/2 \rfloor -1$. We do not adopt this method here since it increases $k_{\textup{eff}}^{\textup{b}}$, increases the maximum magnitude of the running disparity to $m+x$, in addition to its complexity.
\end{remark}

We are now ready to discuss the rate of CB-LOCO codes. A CB-LOCO code $\mathcal{C}_{m,x}^{\textup{cb}}$, with $x$ bridging bits/symbols associated to each codeword, has rate:
\begin{align}\label{eqn_rateb}
R_{\textup{LOCO}}^{\textup{cb}} &= \frac{\left \lfloor \log_2 \left ( \frac{1}{2} N^{\textup{cb}}(m, x) \right )  \right \rfloor}{m+x} \nonumber \\ &= \frac{\left \lfloor \log_2 \left( N(m, x) - 2 \right )  \right \rfloor - 1}{m+x},
\end{align}
where $N(m, x)$ is obtained from the recursive relation (\ref{eqn_rec}). The numerator, which is $\left \lfloor \log_2 \left( N(m, x) - 2 \right )  \right \rfloor - 1$, is the length of the messages $\mathcal{C}_{m,x}^{\textup{cb}}$ encodes.

Comparing the rate of the CB-LOCO code $\mathcal{C}_{m,x}^{\textup{cb}}$ to the C-LOCO code $\mathcal{C}_{m,x}^{\textup{c}}$ via subtracting (\ref{eqn_rateb}) from (\ref{eqn_rate}) gives:
\begin{align}
&R_{\textup{LOCO}}^{\textup{c}} - R_{\textup{LOCO}}^{\textup{cb}} \nonumber \\ &= \frac{\left \lfloor \log_2 \left( N(m, x) - 2 \right )  \right \rfloor}{m+x} - \frac{\left \lfloor \log_2 \left( N(m, x) - 2 \right )  \right \rfloor - 1}{m+x}. \nonumber
\end{align}
Consequently,
\begin{align}\label{eqn_ratediff}
R_{\textup{LOCO}}^{\textup{c}} - R_{\textup{LOCO}}^{\textup{cb}} = \frac{1}{m+x}.
\end{align}

Under the balancing approach of having two codewords to encode each message, the maximum number of codewords corresponding to distinct messages drops to at most half the cardinality of the unbalanced code. Thus, a balanced code achieves the minimum rate loss if the code has a rate loss of only ${1}/{(\textup{code length})}$ with respect to its unbalanced code; since this means the balanced code contains all the codewords of the unbalanced code. In other words, for each codeword in the unbalanced code, there exists another codeword to be paired with, such that the two codewords have their disparities with the same magnitude but opposite signs. Consequently, no codewords are skipped from the unbalanced code in order to achieve balancing. We refer to this rate loss as the \textbf{one-bit minimum penalty} because it can be viewed as a reduction of one bit from the message length. From the above discussion and (\ref{eqn_ratediff}), our CB-LOCO codes achieve the minimum rate loss, i.e., they achieve the one-bit minimum penalty.

Observe that \textbf{asymptotically, i.e., as $m \rightarrow \infty$, the rate loss resulting from balancing LOCO codes tends to zero} from (\ref{eqn_ratediff}). Thus, CB-LOCO codes asymptotically achieve the same rates as C-LOCO codes. Moreover, the penalty of (rate loss due to) balancing LOCO codes has the highest possible vanishing rate with $m$. As shown in Table~\ref{table_7}, the rate of the moderate-length CB-LOCO code $\mathcal{C}_{116,1}^{\textup{cb}}$ (resp., $\mathcal{C}_{120,2}^{\textup{cb}}$) is within only $1.5\%$ (resp., $2\%$) from the capacity of an unbalanced $\mathcal{T}_x$-constrained code having $x=1$ (resp., $x=2$). As far as we know, balancing other constrained codes in the literature always incurs a notable rate loss, even asymptotically, with respect to the unbalanced codes \cite{knuth_bal, braun_lex, qin_flash}, which is not the case for LOCO codes. For example, the balancing penalty in \cite{knuth_bal} is an added redundancy of more than $\log_2 m$ (see also \cite{knuth_mod}), which is a costly penalty. Moreover, in order to reduce the rate loss due to balancing, the authors of \cite{braun_lex} are adopting large code lengths, which is not needed for LOCO codes. In the finite-length regime, we achieve a higher rate at the same code length or the same rate at a smaller code length in comparison with \cite{braun_lex}.

\begin{table}
\caption{Rates and adder sizes of CB-LOCO codes $\mathcal{C}_{m,x}^{\textup{cb}}$ for different values of $m$ and $x$. The unbalanced capacity is $0.6942$ for $x=1$ and $0.5515$ for $x=2$.}
\vspace{-0.5em}
\centering
\scalebox{1.00}
{
\begin{tabular}{|c|c|c|}
\hline
\makecell{Values of $m$ and $x$} & \makecell{$R_{\textup{LOCO}}^{\textup{cb}}$} & \makecell{Adder size} \\
\hline
$m=14$, $x=1$ & $0.6000$ & $9$ bits \\
\hline
$m=24$, $x=1$ & $0.6400$ & $16$ bits \\
\hline
$m=44$, $x=1$ & $0.6667$ & $30$ bits \\
\hline
$m=54$, $x=1$ & $0.6727$ & $37$ bits \\
\hline
$m=80$, $x=1$ & $0.6790$ & $55$ bits \\
\hline
$m=116$, $x=1$ & $0.6838$ & $80$ bits \\
\hline
$m=8$, $x=2$ & $0.4000$ & $4$ bits \\
\hline
$m=15$, $x=2$ & $0.4706$ & $8$ bits \\
\hline
$m=24$, $x=2$ & $0.5000$ & $13$ bits \\
\hline
$m=42$, $x=2$ & $0.5227$ & $23$ bits \\
\hline
$m=73$, $x=2$ & $0.5333$ & $40$ bits \\
\hline
$m=120$, $x=2$ & $0.5410$ & $66$ bits \\
\hline
\end{tabular}}
\label{table_7}
\vspace{-0.5em}
\end{table}

\begin{example}\label{ex_8}
Consider again the B-LOCO code $\mathcal{C}_{6,1}^{\textup{b}}$ in Table~\ref{table_6}. From (\ref{eqn_kbeff}), the CB-LOCO code $\mathcal{C}_{6,1}^{\textup{cb}}$ derived from $\mathcal{C}_{6,1}^{\textup{b}}$ has:
\begin{equation}
k_{\textup{eff}}^{\textup{cb}} = 2(6-1) + 1 = 11. \nonumber
\end{equation}

The length of the messages $\mathcal{C}_{6,1}^{\textup{cb}}$ encodes is:
\begin{equation}
\left \lfloor \log_2 \left( N(6, 1) - 2 \right )  \right \rfloor - 1 = \left \lfloor \log_2 24  \right \rfloor - 1 = 3. \nonumber
\end{equation}
The CB-LOCO code $\mathcal{C}_{6,1}^{\textup{cb}}$ is also shown in Table~\ref{table_6} for all messages. From (\ref{eqn_rateb}), the rate of $\mathcal{C}_{6,1}^{\textup{cb}}$ is:
\begin{equation}
R_{\textup{LOCO}}^{\textup{cb}} = \frac{\left \lfloor \log_2 24  \right \rfloor - 1}{6+1} = \frac{3}{7} = 0.4286. \nonumber
\end{equation}
\end{example}

For bigger values of $m$, the rate of a CB-LOCO code $\mathcal{C}_{m,x}^{\textup{cb}}$ exceeds $0.6667$ (resp., $0.5000$) for $x=1$ (resp., $x=2$) as shown in Table~\ref{table_7} and discussed before Example~\ref{ex_8}. These rates cannot be achieved for practical balanced FSM-based RLL codes having $d=x$. Moreover, even to approach these rates, the encoding-decoding complexity of the balanced FSM-based RLL code will be significantly larger than that of the CB-LOCO code. CB-LOCO codes also offer a better rate-complexity trade-off compared with balanced FSM-based $\mathcal{T}_x$-constrained codes. Recall that the rate of a practical FSM-based unbalanced constrained code is typically $0.6667$ (resp., $0.5000$) for $d=x=1$ (resp., $d=x=2$) \cite{siegel_mr, siegel_const}.

Algorithms~\ref{alg_enc} and \ref{alg_dec} can be modified to encode and decode CB-LOCO codes. The major changes are:
\begin{enumerate}
\item For both algorithms, the message length (adder size) is changed to $s^{\textup{cb}} = \left \lfloor \log_2 \left( N(m, x) - 2 \right )  \right \rfloor - 1$.

\item For Algorithm~\ref{alg_enc}, the message here is encoded to $\bold{c} = \bold{c}^0$ initially. After Step~40, $p(\bold{c}^0)$ is calculated. Then, a check is made on the disparities $p_{\textup{r}}$ and $p(\bold{c}^0)$. If $p_{\textup{r}}$ and~$p(\bold{c}^0)$ have the same sign, the codeword complement of $\bold{c}^0$ is transmitted (written), i.e., $\bold{c} = \bold{c}^1$, and $p(\bold{c}) = p(\bold{c}^1) = -p(\bold{c}^0)$. Otherwise, $\bold{c} = \bold{c}^0$ is transmitted (written), and $p(\bold{c}) = p(\bold{c}^0)$. The updated running disparity $p_{\textup{r}}$ is then calculated for the next codeword using $p_{\textup{r}} \leftarrow p_{\textup{r}} + p(\bold{c})$. Only $p(\bold{c})$ is needed because we use the first bridging method.

\item Let $o(\bold{c})$ be the number of $1$'s in codeword $\bold{c}$ in $\mathcal{C}_{m,x}^{\textup{cb}}$. For Algorithm~\ref{alg_enc}, $p(\bold{c})$ can be easily computed from:
\begin{equation}
p(\bold{c}) = 2o(\bold{c})-m.
\end{equation}

\item For Algorithm~\ref{alg_dec}, Steps~5, 6, and 7 are removed. Moreover, if $c_{m-1}=0$, the condition under which $g^{\textup{b}}(\bold{c})$ is increased by $\frac{1}{2} N(i-x+1, x)$ remains ``\textbf{if} $c_i=1$'' from (\ref{eqn_g20}) in Theorem~\ref{thm_induct2}. However, if $c_{m-1}=1$, the condition under which $g^{\textup{b}}(\bold{c})$ is increased by $\frac{1}{2} N(i-x+1, x)$ becomes ``\textbf{if} $c_i=0$'' from (\ref{eqn_g21}) in Theorem~\ref{thm_induct2}.
\end{enumerate}

Table~\ref{table_7} also links the rate of a CB-LOCO code with its encoding and decoding complexity through the size of the adders to be used.

\begin{remark}
Observe that $(d, \infty)$ LO-RLL codes constructed as shown in \cite{tang_bahl} or via the ideas in Remark~\ref{rmk_lorll} do not have the balancing advantage of LOCO codes, which is the complement rule in Lemma~\ref{lem_inv}. In other words, given a LO-RLL codeword, there does not necessarily exist another LO-RLL codeword such that their disparities have the same magnitude but opposite signs after NRZI signaling. Therefore, balancing these codes is associated with a higher penalty compared with balancing LOCO codes as a result of the many unused codewords. This is another advantage of LOCO codes over $(d, \infty)$ LO-RLL codes in addition to the rate-complexity trade-off advantage illustrated in Remark~\ref{rmk_lorll} and Remark~\ref{rmk_lorll_2}.
\end{remark}

%%%%%%%%%%%%%%%%%%%%%%%%%%%%%%%%%%%%%
\section{Conclusion}\label{sec_conc}

We introduced LOCO codes, a new family of constrained codes, where the combination of recursive structure and lexicographic indexing of codewords enables simple mapping-demapping between the index and the codeword itself. We showed that this mapping-demapping enables low complexity encoding and decoding algorithms. We also showed that LOCO codes are capacity-achieving, and that at moderate lengths, they provide a rate gain of up to $10\%$ compared with other practical constrained codes that are used to achieve the same goals. Inherent symmetry of LOCO codes makes balancing easy. We demonstrated that the rate loss associated with balancing LOCO codes is minimal, and that this loss tends to zero in the limit, so that balanced LOCO codes achieve the same asymptotic rates as their unbalanced counterparts. Moreover, we demonstrated a density gain of about $20\%$ in modern MR systems by using a LOCO code to protect only the parity bits of an LDPC code via mitigating ISI. We suggest that LOCO codes provide a simple and effective practical method for improving the performance of a wide variety of data storage and computer systems. Ongoing work includes asymmetric and non-binary LOCO codes.

%%%%%%%%%%%%%%%%%%%%%%%%%%%%%%%%%%%%%
\section*{Acknowledgment}\label{sec_ack}

The authors would like to thank the associate editor Prof. Anxiao Jiang for handling the paper and for the constructive feedback. The authors would also like to thank the anonymous reviewers for their valuable and helpful comments (this extends to the ITW reviewers as well).

%%%%%%%%%%%%%%%%%%%%%%%%%%%%%%%%%%%%%

\end{document}